%% file: main.tex
\documentclass[conference]{IEEEtran}
\IEEEoverridecommandlockouts
\usepackage{amsmath,amssymb,amsfonts}
\usepackage{algorithmic}
\usepackage{graphicx}
\usepackage{textcomp}
\usepackage[table,xcdraw]{xcolor}
\def\BibTeX{{\rm B\kern-.05em{\sc i\kern-.025em b}\kern-.08em
    T\kern-.1667em\lower.7ex\hbox{E}\kern-.125emX}}
\usepackage{mathptmx} % This is Times font
\usepackage{fancyhdr}
\usepackage[normalem]{ulem}
\usepackage[hyphens]{url}
\usepackage[sort,nocompress]{cite}
\usepackage[final]{microtype}

\usepackage{adjustbox}
\usepackage{enumitem}
\usepackage{booktabs}
\usepackage{xspace}
\usepackage{multirow}
\usepackage{multicol}
\usepackage{setspace}
 
\usepackage[bookmarks=true,breaklinks=true,letterpaper=true,colorlinks,citecolor=blue,linkcolor=blue,urlcolor=blue]{hyperref}

\begin{document}
\newcommand{\sysname}{Darwin\xspace}
\makeatletter
\newcommand{\newlineauthors}{%
  \end{@IEEEauthorhalign}\hfill\mbox{}\par
  \mbox{}\hfill\begin{@IEEEauthorhalign}
}
\makeatother

%%%%%%%%%%%%%%%%%%%%%%%%%%%%%%%%%%%%

%%%%%%%%%%%---SETME-----%%%%%%%%%%%%%
\title{\sysname: A DRAM-based Multi-level Processing-in-Memory Architecture for Data Analytics}
%\thanks{Identify applicable funding agency here. If none, delete this.}} 

\author
{
\IEEEauthorblockN{\large Donghyuk Kim}
\IEEEauthorblockA{\textnormal{kar02040@kaist.ac.kr}\\
\textnormal{KAIST}\\\textnormal{Daejeon, South Korea}}
\and
\IEEEauthorblockN{\large Jae-Young Kim}
\IEEEauthorblockA{\textnormal{jykim1109@kaist.ac.kr}\\
\textnormal{KAIST}\\\textnormal{Daejeon, South Korea}}
\and
\IEEEauthorblockN{\large Wontak Han}
\IEEEauthorblockA{\textnormal{11tak@kaist.ac.kr}\\
\textnormal{KAIST}\\\textnormal{Daejeon, South Korea}}
\newlineauthors
\IEEEauthorblockN{\large Jongsoon Won}
\IEEEauthorblockA{\textnormal{jongsoon.won@sk.com}\\
\textnormal{SK hynix Inc.}\\\textnormal{Icheon, South Korea}}
\and
\IEEEauthorblockN{\large Haerang Choi}
\IEEEauthorblockA{\textnormal{haerang.choi@sk.com}\\
\textnormal{SK hynix Inc.}\\\textnormal{Icheon, South Korea}}
\and
\IEEEauthorblockN{\large Yongkee Kwon}
\IEEEauthorblockA{\textnormal{yongkee.kwon@sk.com}\\
\textnormal{SK hynix Inc.}\\\textnormal{Icheon, South Korea}}
% \newlineauthors
\and
\IEEEauthorblockN{\large Joo-Young Kim}
\IEEEauthorblockA{\textnormal{jooyoung1203@kaist.ac.kr}\\
\textnormal{KAIST}\\\textnormal{Daejeon, South Korea}}
}

\maketitle

%%%%%% -- PAPER CONTENT STARTS-- %%%%%%%%

\input{Contents/0_Abstract}
% \vspace{0.5in}
%\setstretch{0.95}
\input{Contents/1_Intro}
\input{Contents/2_Background}

\input{Contents/3_Motivation}

\input{Contents/4_DBDRAMPIM}
\input{Contents/5_InmemoryLogicDesign}

\input{Contents/6_SoftwareStack}

\input{Contents/7_Methodology}

\input{Contents/8_ExperimentalResult}

\input{Contents/9_RelatedWork}
\input{Contents/10_Conclusion}

%%%%%%%%% -- BIB STYLE AND FILE -- %%%%%%%%
\bibliographystyle{IEEEtranS}
\bibliography{refs}
%%%%%%%%%%%%%%%%%%%%%%%%%%%%%%%%%%%%

\end{document}

%% file: Contents/0_Abstract.tex
\begin{abstract}

Processing-in-memory (PIM) architecture is an inherent match for data analytics application, but we observe major challenges to address when accelerating it using PIM. First, data analytics involves intensive read and write operations on databases, causing bandwidth bottleneck issues even inside the memory. Furthermore, irregular and non-deterministic data analytics workload causes load imbalance among in-memory processing units, deteriorating the overall performance. Then, the conventional DRAM command protocol, which sends a command to a single bank, causes the command bottleneck for complex data analytics operators. In this paper, we propose \sysname, a practical LRDIMM-based multi-level PIM architecture for data analytics, which fully exploits the internal bandwidth of DRAM using the bank-, bank group-, chip-, and rank-level parallelisms. Considering the properties of data analytics operators and DRAM’s area constraints, \sysname maximizes the internal data bandwidth by placing the PIM processing units, buffers, and control circuits across the hierarchy of DRAM. More specifically, it introduces the bank processing unit for each bank in which a single instruction multiple data (SIMD) unit handles regular data analytics operators (e.g., select, aggregate, and sort), and bank group processing unit for each bank group to handle workload imbalance in the condition-oriented data analytics operators (e.g., project, and join). Furthermore, \sysname supports a novel PIM instruction architecture that concatenates instructions for multiple thread executions on bank group processing entities, addressing the command bottleneck by enabling separate control of up to 512 different in-memory processing units simultaneously. 
We build a cycle-accurate simulation framework to evaluate \sysname with various DRAM configurations, optimization schemes and workloads. \sysname achieves up to 14.7x speedup over the non-optimized version, leveraging many optimization schemes including the novel instruction architecture, circuit optimizations for each data analytics operator, and the proper placement of the processing unit. Finally, the proposed \sysname architecture achieves 4.0x-43.9x higher throughput and reduces energy consumption by 85.7\% than the baseline CPU system (Intel Xeon Gold 6226 + 4 channels of DDR4-2933) for essential data analytics operators. Compared to the state-of-the-art PIM architectures, \sysname achieves up to 7.5x and 7.1x in the basic query operators and TPC-H queries, respectively. \sysname is based on the latest GDDR6 and requires only 5.6\% area overhead, suggesting a promising PIM solution for the future main memory system.

\end{abstract}

%% file: Contents/1_Intro.tex
\section{Introduction}
\label{introduction}

In the era of big data, data-intensive applications such as artificial intelligence~\cite{lecun2015deep} and data analytics~\cite{chaudhuri1997overview} proliferate by utilizing extremely large datasets. As these applications mainly consist of low compute-to-memory ratio operations, memory operations dominate over the compute operations, causing the “memory wall”~\cite{boroumand2018google}. 
%As a result, extensive data frequently travels through narrow off-chip bandwidth from storage to CPU, causing the “memory wall”~\cite{boroumand2018google} in conventional von Neumann architecture. 
Furthermore, the bottleneck on the memory side exacerbates as the improvement of memory technology in speed falls behind the logic technology. Continual efforts in increasing the off-chip bandwidth of recent DRAM technologies \cite{lee201425, hwang201816gb} result in higher IO speed and more pins, but they come at the cost of expense and power consumption.

To minimize the data movement overhead in data analytics, the database backend is relocated from storage to main memory \cite{stonebraker2013voltdb,lahiri2013oracle,lindstrom2013ibm}, avoiding expensive disk IO access. 
%which causes larger energy consumption and longer latency. 
Additionally, some of analytical query operators (e.g., \textit{select}, \textit{aggregate}, \textit{sort}, \textit{project}, and \textit{join}) are converted into vector operations, increasing the throughput of the query processing \cite{boncz2005monetdb}. Since the query operators iteratively compute on sequences of data stream, vector type processing can easily accelerate by computing many data at a time. However, even with the high-performance CPU, the vectorized query operations with low compute-to-memory ratio cannot be sped up due to the ever-growing data size and limitation of the off-chip bandwidth. 

Previous research propose accelerating data analytics on different hardware platforms, such as field-programmable gate array (FPGA)~\cite{owaida2017centaur, watanabe2019column, xu2020aquoman} and graphics processing unit (GPU)~\cite{li2016hippogriffdb, shanbhag2020study}. However, these approaches focus on improving the computation capability, while leaving the essence of the memory bottleneck issue that occurs in computing data-intensive applications. Therefore, the paradigm shift from computation-centric to memory-centric architecture is unavoidable in such data-intensive applications.

As a result, inevitable memory bottleneck problem drives both industry and academia to reassess the DRAM-based near-memory-processing (NMP)~\cite{drumond2017mondrian, xu2020aquoman, boroumand2021polynesia,kwon2019tensordimm, ke2020recnmp} and processing\textendash in\textendash memory (PIM)~\cite{li2017drisa, he2020newton, gu2020ipim, lee2021hardware, kim2022overview, lenjani2020fulcrum, lenjani2022gearbox, xie2021spacea, gao2019computedram, kim2021gradpim, xin2020elp2im, park2021trim} architectures that increase the internal bandwidth by integrating computational logic and DRAM device/cells closely. NMP architectures integrate homogeneous processing unit (PU) per vault in the base logic die of hybrid memory cube (HMC), supporting flexible dataflow for query operations. % processing unit in the vault can independently control each memory stack, which guarantees flexible dataflow in computing complicated query plans. 
However, these approaches, only integrating PUs external to memory, lose the opportunity to fully benefit from the wide internal bandwidth of DRAM memory. On the other hand, PIM architectures can fully exploit the abundant internal bandwidth of DRAM. However, they cannot efficiently compute complex query operations with complicated internal data movement since they are only capable of processing bulk-wise data processing such as vector-vector and matrix-vector multiplications with fixed data path. Furthermore, PIM architectures that integrate compute logic closer than the bank level (e.g., in cell and near subarray) are impractical that reduces the density of cells significantly.

\begin{figure}[t]
\centering
\includegraphics[width=\columnwidth]{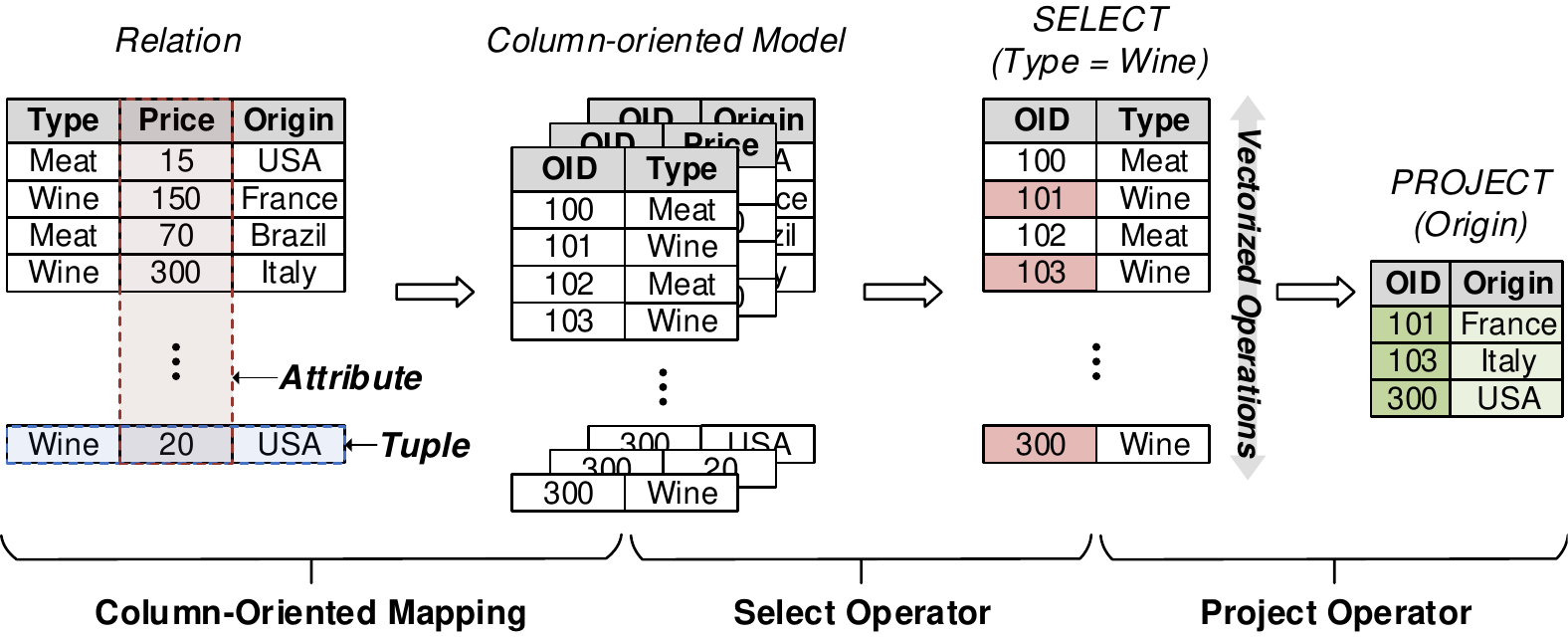}
\caption{Column-oriented Database Management System}
\label{Figure_ColumnDB}
\end{figure}

To address the limitations of the previous approaches, we propose \sysname, a practical LRDIMM-based multi-level PIM architecture for data analytics.
First, \sysname reuses the conventional DRAM hierarchical architecture to save the additional interconnect resource for internal data movement occurred for PIM computation. It exploits bank, bank group, and rank levels for multi-level parallelisms with reduced internal data movement. It utilizes single instruction multiple data (SIMD) fixed-point unit exploiting wide bank-level parallelism while separating the control granularity to bank group for flexible execution. 

Second, \sysname supports in-memory control unit to support seamless workload balancing after condition-oriented query operators where the output data size is not predictable in static compile time. 
Third, \sysname modifies the command interface to avoid the command bottleneck in individually controlling multiple PUs simultaneously. We introduce a new PIM instruction architecture that concatenates multi-bank group command that enables independent but concurrent operations in multiple bank groups.

In this paper, we make the following contributions:
\begin{itemize}
  \item We propose a multi-level PIM architecture for data analytics which fully exploits the internal bandwidth from the bank, bank group, and rank levels within a commodity DRAM architecture, reducing any additional overhead for practicality.
  \item We propose bank group-level processing units to support irregular data analytics operations, enabling dynamic runtime execution on the condition-oriented operations and low-overhead workload balancing.
  \item We propose a new command interface to support parallel but individual executions of in-memory PUs while avoiding the command bottleneck. 
  \item We evaluate TPC-H and basic query operators on \sysname over the baseline CPU and state-of-the-art PIM.
\end{itemize}

%% file: Contents/2_Background.tex
\section{Background}
\label{sec_background}

\subsection{Data Analytics}
Most current database systems used in finance and business are relational database management system (RDBMS) \cite{codd1990relational}, where data are stored in the form of relations, which comprise lists of tuples and attributes. A tuple represents a row, and an attribute represents a column in a relation. RDBMS can be divided into two types depending on how data are stored: row-oriented \cite{armbrust2015spark} and column-oriented \cite{stonebraker2005c, idreos2012monetdb}. The row-oriented database organizes data by the record, sequentially storing attributes of records. It is optimized for reading and writing rows, such as online transactional processing (OLTP). On the other hand, a column-oriented database such as MonetDB \cite{idreos2012monetdb} stores data into an array of each attribute, which benefits in reading and computing on columns as in online analytical processing (OLAP), as shown in Figure~\ref{Figure_ColumnDB}. Furthermore, the basic operators can turn into vectorized query executions by using column-oriented storage, where the execution is iteratively performed in a batch of input data. As a result, sequential memory accesses are prevalent in column-oriented database, where PIM are suitable as it promotes high data parallelism and wide bandwidth utilization of the memory.

In the column-oriented storage, \textit{project} is the dominant operator among the others. Kepe et al.~\cite{kepe2019database} analyze the latency breakdown of MonetDB in the TPC-H benchmark. The result shows that the \textit{project} takes up 58\% of the overall latency, while each of the other operator (e.g., \textit{select}, \textit{aggregate}, \textit{sort}, and \textit{join}) takes only up to 11\%. The \textit{project}, which materializes intermediate tables, occupies the majority because it occurs after every query operator in a query plan. This is because the column-oriented storage generates sets of object-IDs (OIDs), representing the address of each tuple which is unique for each relation, as the output of query operators. Using the OID result, the \textit{project} can connect the previous and following operators by generating the intermediate tables, which is used in the following operator as inputs.

\subsection{Architecture of Main Memory System}
In order to properly exploit the PIM architecture, we need to understand the control granularity and the internal bandwidth of main memory system. The internal bandwidth can be understood through the logical structure of the main memory. It adopts the multi-drop tree topology, where the highest level starts from the memory channel controlled by a memory controller of the host. A channel comprises multiple dual in-line memory modules (DIMMs). Within a DIMM, several DDR chip packages are placed forming a rank. The number of chip packages in a rank is determined by the number of DQ pins per package, where DQ pins are used for data input and output. The total number of DQ pins per DIMM is 64 bits to match the JEDEC specification. Then, multiple bank groups make up a rank where four banks form a bank group. Due to the multi-drop tree topology, only a single bank of the DDR packages within a rank can be accessed at a time across a channel. Getting the same command and address from the memory controller, each DDR package in a rank operates simultaneously, contributing to a part of the combined DQ pins. Even though DRAM can only access a bank at a time, it is efficient for DRAM since it can exploit the bank interleaving scheme to hide the internal delays, such as activation and precharge delays, increasing the bandwidth utilization. However, the multi-drop topology unavoidably limits the internal bandwidth of multiple banks for  simultaneous read/write.

%% file: Contents/3_Motivation.tex
\section{Challenge of Data Analytics}
\label{motivation}

\subsection{Internal Data Movement Overhead in Single\textendash Level PIM}
\label{subsection_motiv_1}

\begin{figure}
\centering
\includegraphics[width=\columnwidth]{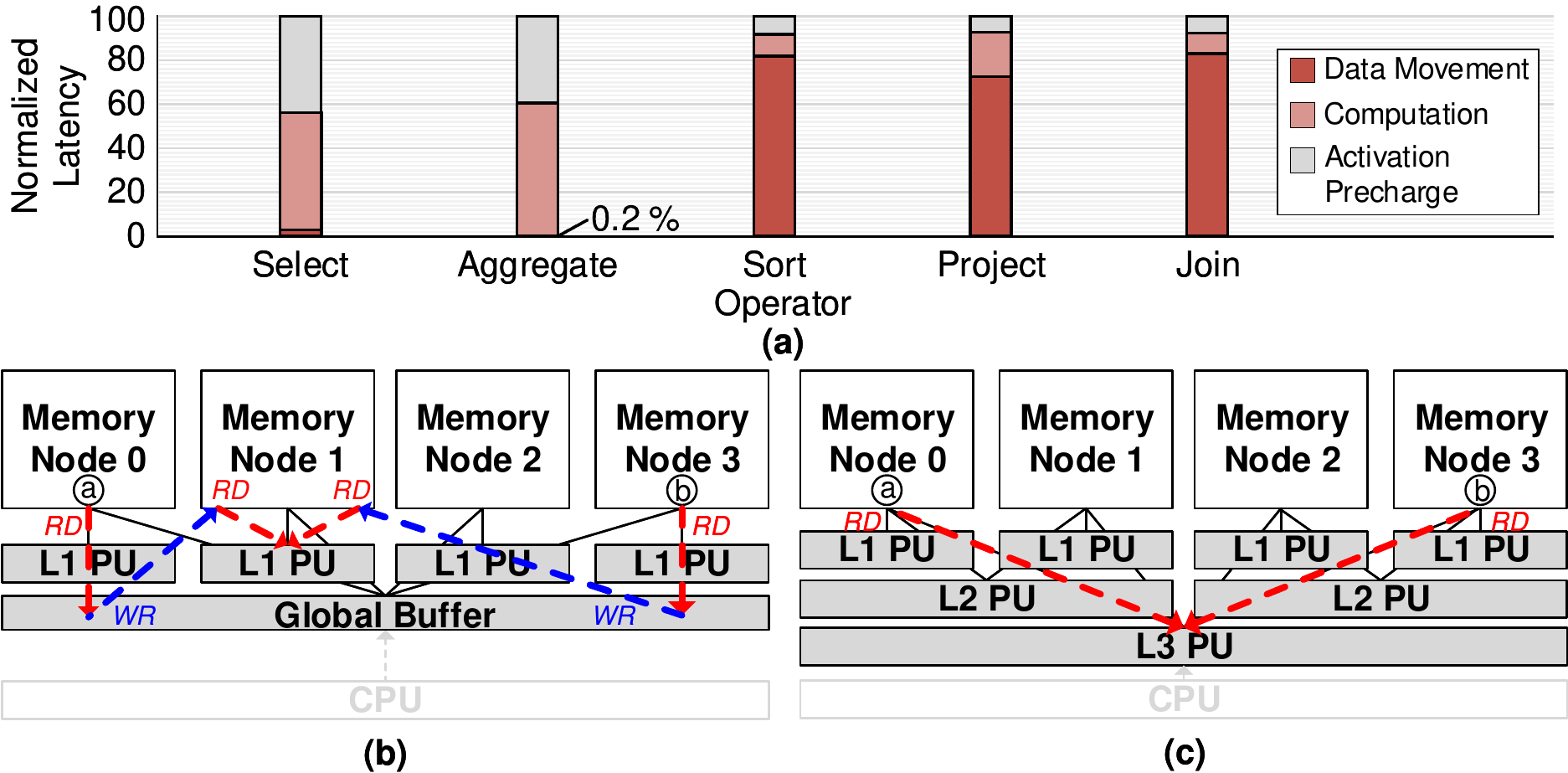}
\caption{(a) Latency Breakdown of Data Analytics Operators (b) Single-Level PIM (c) Multi-Level PIM.}
\label{Figure_0}
\end{figure}

The conventional single-level PIM (SLPIM) architectures, such as proposed in~\cite{drumond2017mondrian, boroumand2021polynesia, he2020newton, lee2021hardware},  incur significant internal data movement in accelerating condition-oriented operators (e.g., \textit{join} and \textit{project}) and merge phase of \textit{sort}. SLPIM refers to an architecture that places PUs only at a single level (e.g., subarray, bank, or rank). We demonstrated to see the overhead of internal data movement in SLPIM when executing the basic data analytics operators. Figure~\ref{Figure_0} (a) shows the latency breakdown of the basic operators executed in SLPIM. The result shows two distinct trends since the characteristics of the operators is  different. First, the execution of \textit{sort}, \textit{project}, and \textit{join} are dominated by the internal data movement. As \textit{project} and \textit{join} are condition-oriented operators, they generate workload imbalance across the PUs and induce additional internal data movement. When executing these operators, dataflow and input/output sizes are decided at runtime. In other words, the input can be evenly mapped on different memory nodes for the balanced workload, but the intermediate data are distributed unevenly as the different outputs are generated among the PUs. It leads to underutilization and performance degradation, especially in the parallel computing of the in-memory PUs. Thus, it requires balancing the workload among the PUs to maximize hardware utilization and performance, which additionally generates data movement. Furthermore, the merge phases of \textit{sort} and \textit{join} cause significant internal data movement because data in different nodes are accessed frequently to merge separate partitions into one. Second, the data movement are negligible in \textit{select} and \textit{aggregate} as they are not condition-oriented nor have heavy merge phases. Instead, the main overhead is computation where they require high computation throughput for vectorized query execution. 

As shown in Figure~\ref{Figure_0} (b), SLPIM's data access mechanism is inherently inefficient when accessing PU's neighbor memory node. SLPIM's data access is degraded since each PU shares global buffer for inter-node data movement which causes the bottleneck in the global buffer. Since data analytics incurs significant overhead in internal data movement, PIM architecture should be capable of moving data inside DRAM efficiently. However, the conventional DRAM structure does not have a specific interconnect for internal data movement. While Rowclone~\cite{seshadri2013rowclone} proposes bulk data copy of a row data across different banks, significant data movement induced in data analytics are flexible that Rowclone cannot be effectively utilized. TransPIM~\cite{zhou2022transpim} and GearBox~\cite{lenjani2022gearbox} propose specific network-on-chip (NoC) for efficient DRAM internal data movement for target applications. However, their NoCs consume a large area overhead considering the DRAM area constraint. In particular, utilizing the customized NoC within a commodity DRAM is unscalable and impractical for the flexible data movement required by data analytics.

Depending on target operations, SLPIM can place PUs at different level (e.g., subarray, bank, or rank). Having a PU at a high level, such as rank~\cite{drumond2017mondrian, boroumand2021polynesia}, enables the direct use of data in multiple memory nodes. However, the advantage of PIM is decreased because the high data parallelism cannot be achieved by placing PU far from the memory nodes. Conversely, placing a PU at a lower level~\cite{he2020newton,lee2021hardware} (e.g., bank) can maximize the advantages of PIM, but the performance is degraded by the frequent internal and off-chip data movement overhead caused by the data analytics operator. When accelerating data analytics, SLPIM is not the most practical architecture since flexibility in data movement and high computation parallelism can not be achieved at the same time.

\subsection{Command Bottleneck}
\label{subsection_motiv_3}

\begin{figure*}
\centering
\includegraphics[width=2\columnwidth]{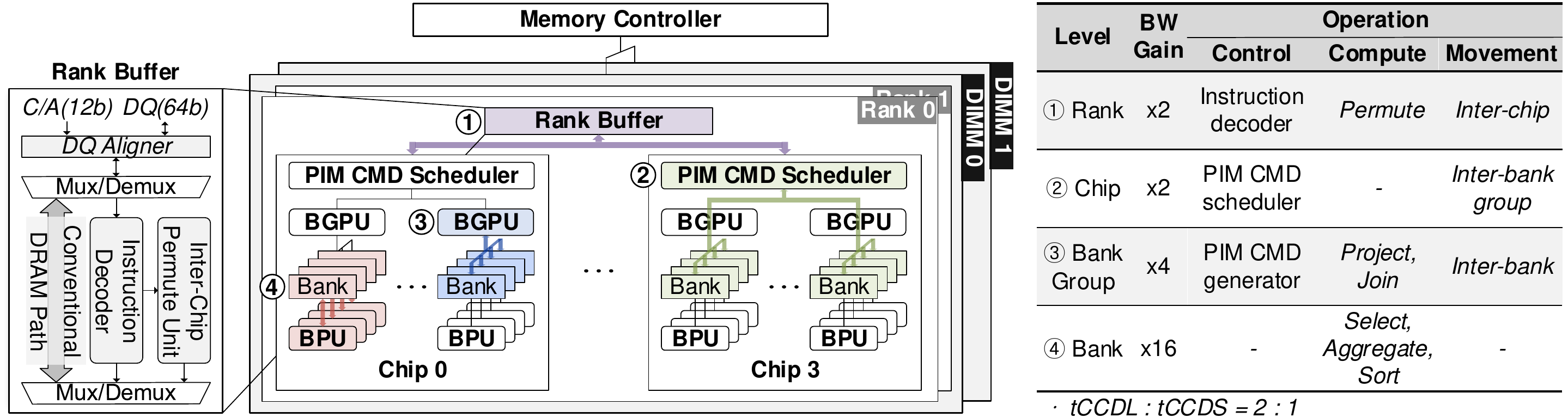}
\caption{Multi-level Architecture of \sysname.}
\label{Figure_ArchitectureOverview}
\end{figure*}

The conventional DRAM command protocol causes a bottleneck in PIM since it is only dedicated to utilize off-chip bandwidth efficiently. It exploits bank interleaving by alternatively accessing data from different banks since it can only send a command (e.g., activation, read, write, precharge, and refresh) to a single bank at a time. It is not suitable for executing PIM operations involving more than one PUs across multiple banks. Regardless of how fast DRAM receives the commands, the shortest latency is saturated at ${t_{CCDS}}$, the minimum interval time between the memory column read. 

To address the command bottleneck, previous research \cite{he2020newton, lee2021hardware} proposed all bank mode. Instead of controlling a single bank, it sends a command to control every bank with the same command. As a result, it can efficiently address the command bottleneck for matrix-vector multiplications that only requires a homogeneous dataflow. However, data analytics incurs irregularity across the PUs requiring different dataflow due to the condition-oriented workload. Thus, all bank mode is not suitable for irregular query operators due to the low control granularity and poor flexibility in which each PU needs to compute the different workload. For example, when processing \textit{join} and \textit{project} in a PIM with multiple PUs, each PU computes a different workload depending on the partitioned input data. Since the required dataflow varies depending on the input data, all bank mode, which is only efficient for applications with homogenous dataflow, is not suitable for the PIM architecture that targets data analytics.

%% file: Contents/4_DBDRAMPIM.tex
%\vspace{-0.04in}
\section{Darwin Architecture}
\label{proposed_architecture}
%\vspace{-0.01in}
 
We propose \sysname, a practical multi-level PIM (MLPIM) architecture with the concatenated instructions, multiple threads (CIMT) to address the challenges of in-memory data analytics processing. \sysname is capable of handling the complex dataflow and imbalanced workload problems, with keeping conventional DRAM's hierarchical structure for practicality. 
%of rank, bank group, and bank to handle the complex dataflow of query operators and mitigate their imbalanced workload overhead. 
Regarding the PIM-host interface, CIMT addresses the command bottleneck for the irregular operators, maximizing the command density and the hardware utilization. %Furthermore, a column-wise relation mapping scheme is adopted to exploit the sequential access pattern and data parallelism in query processing.

%\vspace{-0.05in}
\subsection{Multi-Level PIM Architecture}
%\vspace{-0.01in}

As explained in Section~\ref{subsection_motiv_1}, SLPIM is ineffective in accelerating data analytics. To this end, \sysname integrates heterogeneous processing units at different levels of the DRAM hierarchy to achieve flexible internal data movement. Although this approach seems to increase hardware and software complexity, MLPIM is a practical solution to reduce the complexity of hardware for data analytics which requires complicated dataflow. First,  by integrating hardware units at both high and low levels of DRAM, \sysname reduces hardware overhead and eliminates additional NoC costs. This is achieved by reusing the conventional DRAM network and leveraging efficient memory access across multiple memory nodes. Furthermore, to enable PIM feasible, \sysname integrates optimized PUs that are located no closer than the bank-level. Second, \sysname reduces software complexity by integrating a hardware controller inside DRAM to manage the parallel processing of irregular operators.

As shown in Figure~\ref{Figure_0} (c), the major difference of SLPIM and MLPIM is that the levels where the operators are offloaded make different memory access. Therefore, the levels of PUs must be placed depending on the data access patterns of the operators. When performing regular operators (e.g., \textit{aggregate}, \textit{select}, and \textit{sort}), they are handled most effectively when processed at L1, where the highest bandwidth gain is guaranteed. Since their output data size is determined at the static time, they do not incur workload imbalance. %Note that the output of \textit{select} is bitmask data. 
In addition, each PU requires few memory accesses from its neighbor memory nodes. 
%Therefore, it is handled most effectively when processed at L1, where the highest bandwidth gain is guaranteed.
On the other hand, irregular operators (e.g., \textit{join}, and \textit{project}) are handled most effectively when processed at a higher level where the efficient irregular memory access can be provided. These operators cause frequent irregular memory access to several nodes by a PU. If the PU is placed at a lower level the memory accesses slow down. Furthermore, the workload imbalance occurred by these and merge operators generate additional data movement across memory nodes for balancing the workload.

The MLPIM architecture of Darwin, as depicted in Figure~\ref{Figure_ArchitectureOverview}, incorporates computing units and control across different levels, such as rank, chip, bank group, and bank. The bandwidth gain and corresponding operations at each level are summarized in the table. The major PUs are placed at the bank and bank group levels, with the bank processing unit (BPU) supporting regular operators to maximize data parallelism, while the bank group processing unit (BGPU) handles irregular operators that require data from broad memory nodes by addressing the internal data movement across the banks within a bank group. To efficiently use all compute  units, Darwin has a PIM command scheduler at the chip level that supports bank group-level threading. Each bank group acts as an independent processing entity that executes a thread together, and multiple processing entities can execute multiple threads simultaneously. As depicted in the figure, Darwin supports multi-level data movement within a rank at each level, including inter-bank, inter-bank group, and inter-chip communication.

\begin{figure}
\centering
\includegraphics[width=\columnwidth]{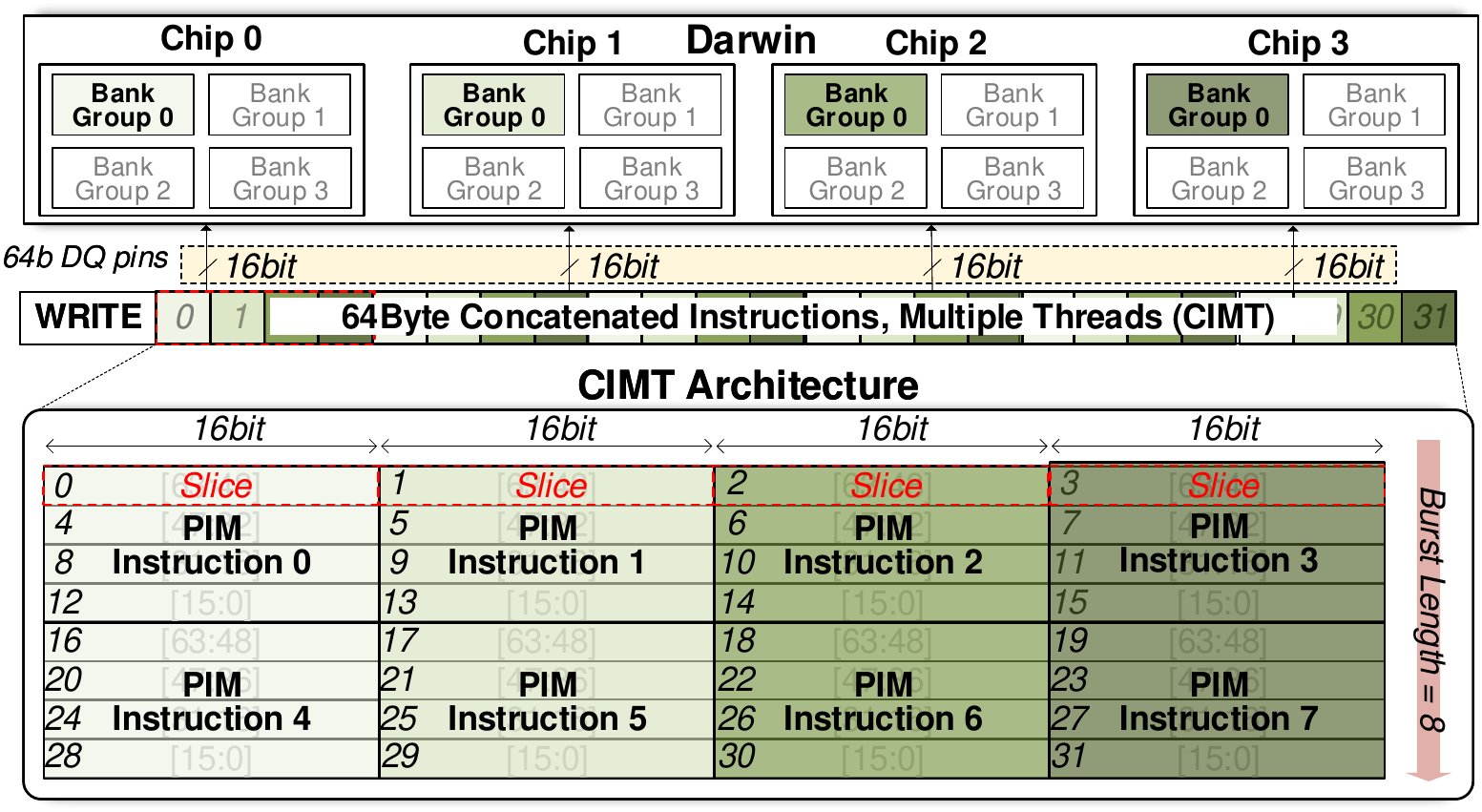}
\caption{Concatenated Instructions, Multiple Threads Architecture}
\label{Figure_CIMT}
\end{figure}

%\vspace{-0.02in}
\subsection{CIMT: Concatenated Instructions, Multiple Threads}
\label{concatenated pim instruction for high control granularity and high data parallelism}
%\vspace{-0.05in}

It is challenging to offload data analytics workload that comprises irregular and regular operators to PIM as stated in Section~\ref{subsection_motiv_3}. To control in-memory PUs separately in computing irregular operators, the conventional DRAM command protocol, which sends one command at a time using the narrow command and address (C/A) pins, cannot provide enough bandwidth to control \sysname. All bank mode can provide wide bandwidth to execute in-memory PUs simultaneously, but the control granularity is extremely low that it cannot process complicated data analytics operations efficiently. To this end, \sysname supports CIMT that handles multiple in-memory PUs with fine control granularity without the command bottleneck. CIMT is optimized for the physical layout of the main memory system that has multiple DRAM chips in a rank. Unlike the conventional command protocol in which different DRAM chips have to receive the same command, each bank group in the different DRAM chips receives different instructions. \sysname utilizes 64-bit DQ pins when sending CIMT using the write command for wider bandwidth. Thus the timing constraints follow the same as the write command.

Each of the four DRAM chips in Figure~\ref{Figure_CIMT} has 16-bit DQ pins, forming a rank with the 64-bit DQ pins for off-chip data transmission. With the burst length of 8, the total data transaction size of 64-byte is sent per a write command. To match the data size, the CIMT instruction comprises 8 different 64-bit PIM instructions being concatenated. Each PIM instruction is divided into four 16-bit slices, and the 16-bit slices of the different PIM instructions are placed in an interleaved manner forming eight 64-bit interleaved instructions. Then, 64-bit interleaved instructions are streamed into \sysname through the 64-bit DQ pins as it is formatted to send the corresponding instruction to each chip. As a result, each chip receives a complete 64-bit instruction. It takes eight cycles to send all 8 PIM instructions with a burst length of 8. 

Each PIM instruction is decoded at the bank group level to generate up to 64 sequential PIM commands to relieve the burden of sending commands through the off-chip bandwidth. The PIM command, generated from the PIM instruction, is a DRAM readable command (e.g., activate, precharge, read, and write) paired with control signals for in-memory PUs. To increase the throughput of BPUs, all banks within the same bank group receive the same PIM command simultaneously, enabling concurrent execution of BPUs. Thus, the CIMT instruction architecture enables separate control of up to 512 different BGPUs simultaneously. The CIMT architecture is applicable to other DRAM configurations.

\vspace{-0.05in}
\subsection{\sysname Operation}
\label{cimt operation example}
\begin{figure}[t]
\centering
\includegraphics[width=\columnwidth]{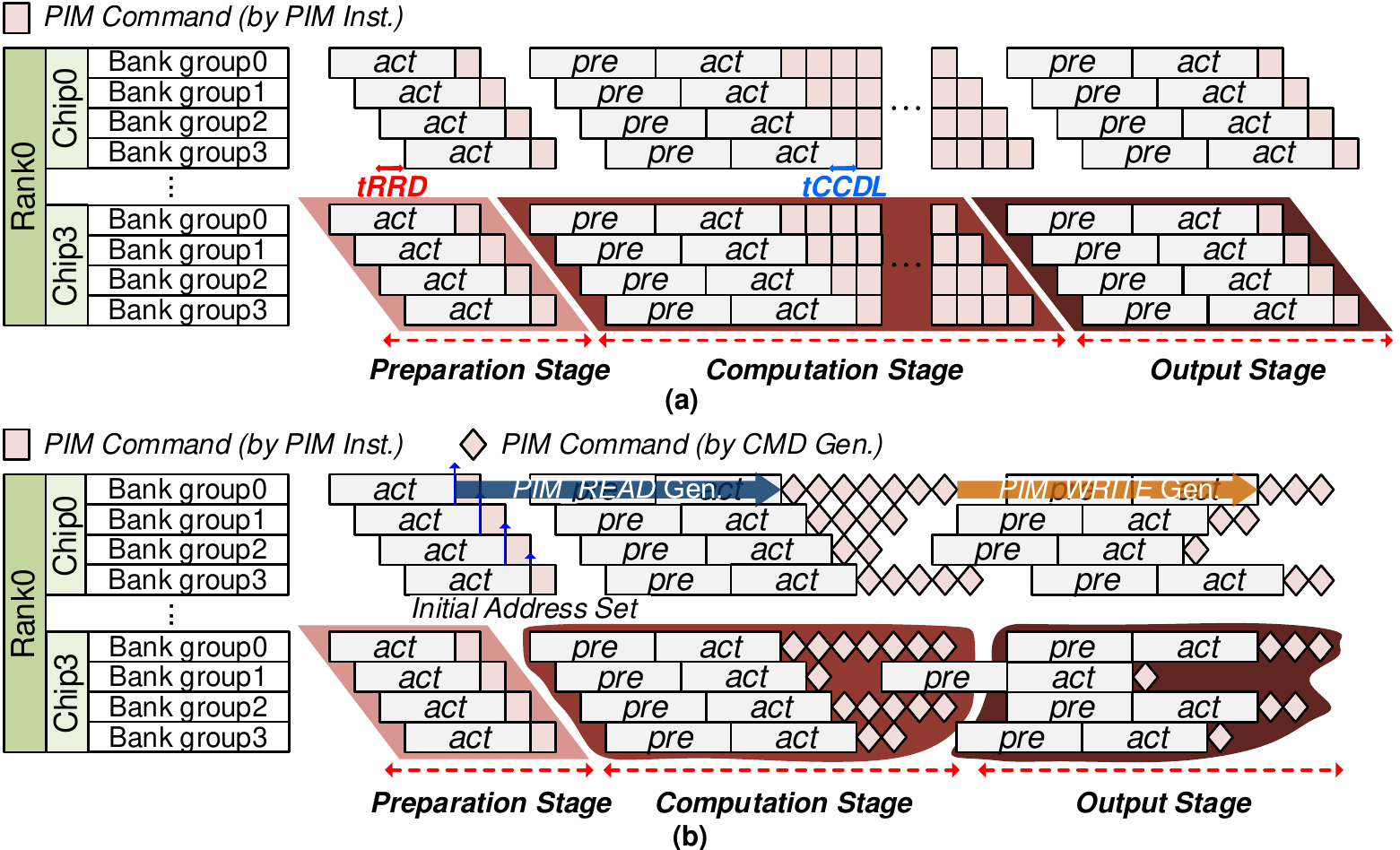}
\caption{\sysname's PIM Operation. (a) Regular Operation (b) Irregular Operation}
\label{Figure_PIM_operation}
\end{figure}

%\vspace{-0.05in}
%\sysname can perform data analytics operators leveraging the multi-level in-memory PUs and its CIMT instruction architecture that addresses command bottleneck. 
Figure~\ref{Figure_PIM_operation} shows \sysname's overall operation flow that is mainly divided into data preparation, computation, and output stages. The regular operators have a fixed data flow in computing vectorized operations. When each thread has the same input amount, the computation flow are equal in all threads. An example operation flow of \textit{select} is shown in Figure \ref{Figure_PIM_operation} (a). In the data preparation stage, the PIM instructions are sent separately to each bank group, and the required input data are transferred to BPU in each bank. To reduce the latency, BPU can directly use the data for one input operand from its own memory. The scalar data of a SIMD operand is sent only once in the preparation stage since the attribute data of the other operand can be transferred directly from memory during the computation stage. In the computation stage, BPUs execute SIMD in parallel. Only one PIM instruction is required to compute the \textit{select} operator on a row of data since 64 sequential PIM commands can be generated. The computation stage continues until the register is filled with the generated output. Having a 512-bit size bitmask register, BPU can compute \textit{select} on one row, which generates a 512-bit output bitmask, and move on to the output stage. In the output stage, the generated output data are stored back in memory. Due to the limited size of the registers in BPU, the output data cannot be stored in the register for the entire operation. For the \textit{select} operator, four write commands are required per a row for 512 bits bitmask data to memory.

The irregular operators have a much more complicated data flow and imbalanced workload among the threads even with the same input amount. An example operation flow of the \textit{project} operator is shown in Figure \ref{Figure_PIM_operation} (b). In the preparation stage, the tuple number and the initial OID are set initially. Then, the 512 bits bitmask data, generated from the previous \textit{select} operator, are transferred to the BGPU. Then, the BGPU receives the PIM instruction to generate corresponding read commands based on the bitmask data for the input attributes. In the computation stage, each BGPU receives different amounts of workload due to the different bitmask that each has. By generating commands internally with the CIMT architecture, each BGPU executes an individual command without the command bottleneck. Furthermore, the computation of the BGPU is rate-matched to the peak bandwidth of the bank group for the streaming execution flow. In the output stage, the selected data are stored in the output register of the BGPU. Once the register is ready, the write commands are generated to store back the output attribute. To balance the workload seamlessly, the BGPU generates the write commands in a bank interleaving manner to write data on each bank memory evenly. It guarantees the shortest latency between the write commands for maximum bandwidth utilization while exploiting the bank group-level parallelism. This process is repeated until the end.

%% file: Contents/5_InmemoryLogicDesign.tex
\section{In-Memory Logic Design}
\subsection{Bank-Level Processing Units}

\begin{figure}[t]
\centering
\includegraphics[width=\columnwidth]{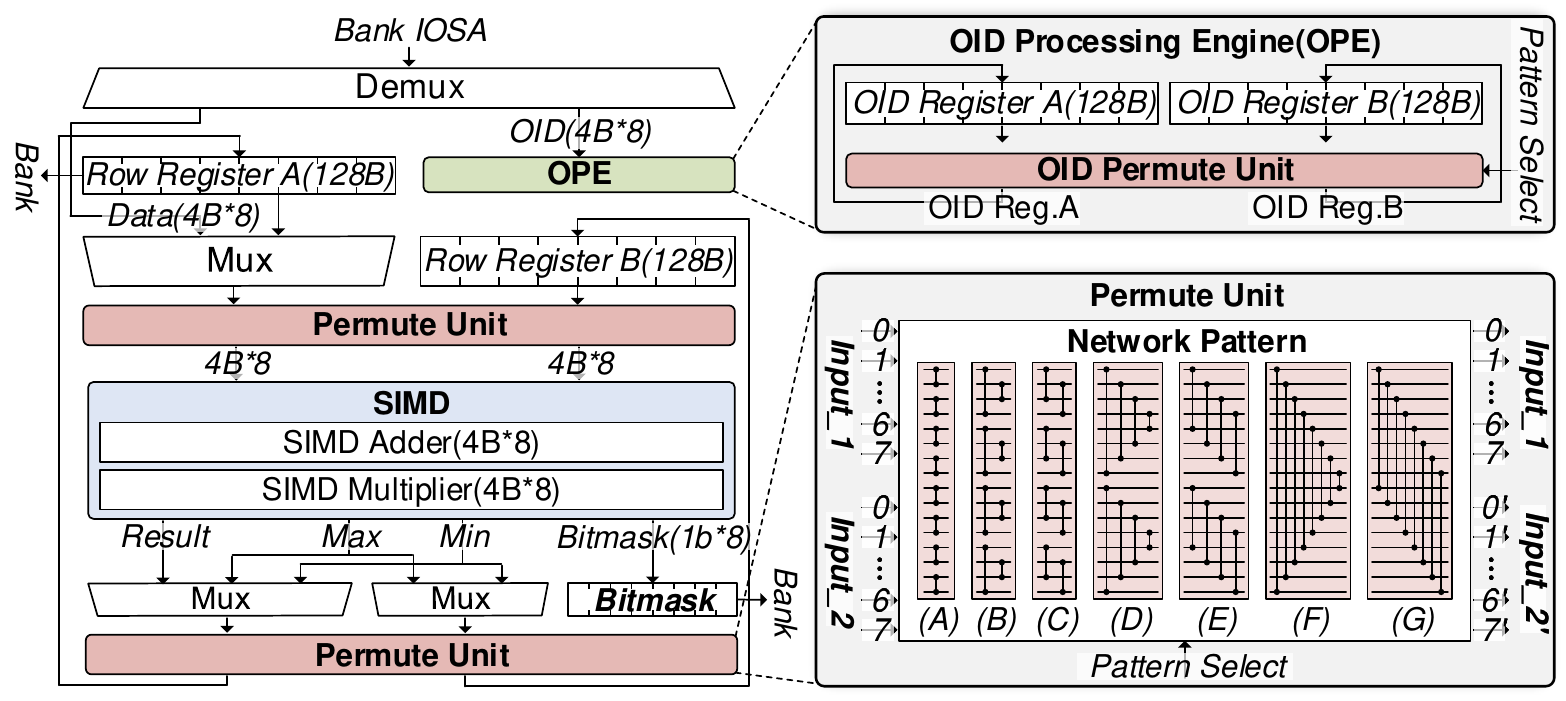}
\caption{BPU Microarchitecture}
\label{Figure_microarchitecture_BPU}
\end{figure}

Figure \ref{Figure_microarchitecture_BPU} shows the BPU's microarchitecture specialized for processing regular operators. Different from the previous PIM architectures designed for matrix-vector multiplication~\cite{he2020newton, lee2021hardware} with straightforward data read and accumulation path, \sysname supports a long sequence of data processing for data analytics, including data read, sort, select, and aggregate. To make this in a streaming processing without data re-writing, the BPU is composed of row registers, a SIMD unit, two permute units, and OID processing engine (OPE). 

\textbf{Select and Aggregate}
The BPU receives 32B of attribute data from the bank's I/O sense amplifiers (IOSAs) and saves in the row register $A$ or $B$. The SIMD unit comprises eight sets of 4-byte fixed-point adders and multipliers, supporting the addition, multiplication, min, and max on eight 4-byte data, which matches the bandwidth of a bank. The SIMD unit outputs bitmask, max, min, and result, and based on the operator's opcode, they are multiplexed into the permute unit. For the \textit{aggregate} operator, the result is accumulated in the row register $A$. For the \textit{select} operator, the bitmask is used in which each 1-bit indicates whether the input tuple is selected or not. Instead of using the 32-bit OID as an output, the bitmask reduces 32x the memory footprint for the output data. The output bitmask is saved in the bitmask register for the selected data, which later can be used for the \textit{project} operator.

\textbf{Sort} We utilize the Bitonic merge-sort algorithm~\cite{batcher1968sorting} to accelerate the rather compute-intensive sort operator with the help of BPU, which is known to work well with SIMD hardware. The Bitonic sort network comprises ten stages that require a total of 4 instructions for each stage, including input permutation, min, max, and output permutation instructions~\cite{chhugani2008efficient}. In addition, the addresses of the data (i.e., OID) must be sorted as well, which incurs doubling the instructions. To minimize computation latency and the number of instructions, we have incorporated two permute units before and after the SIMD unit. The permute unit shuffles sixteen 4B data as input with pre-defined patterns, reducing area overhead by optimizing the permute unit circuitry for seven permutation patterns, as shown in Figure~\ref{Figure_microarchitecture_BPU}. The output generated by the permute unit is sent to the SIMD unit for comparison operations. The SIMD unit generates both min and max data simultaneously, reducing the two instructions for min and max operations. The 16 output data are then sent to the permute unit for output permutation. The BPU supports the OPE, which performs the permutation of addresses that are tagged along with the data result. It eliminates the need to shuffle the OIDs separately as OIDs are shuffled simultaneously with the data. 

\subsection{Bank Group-Level Processing Unit}

\begin{figure}[t]
\centering
\includegraphics[width=\columnwidth]{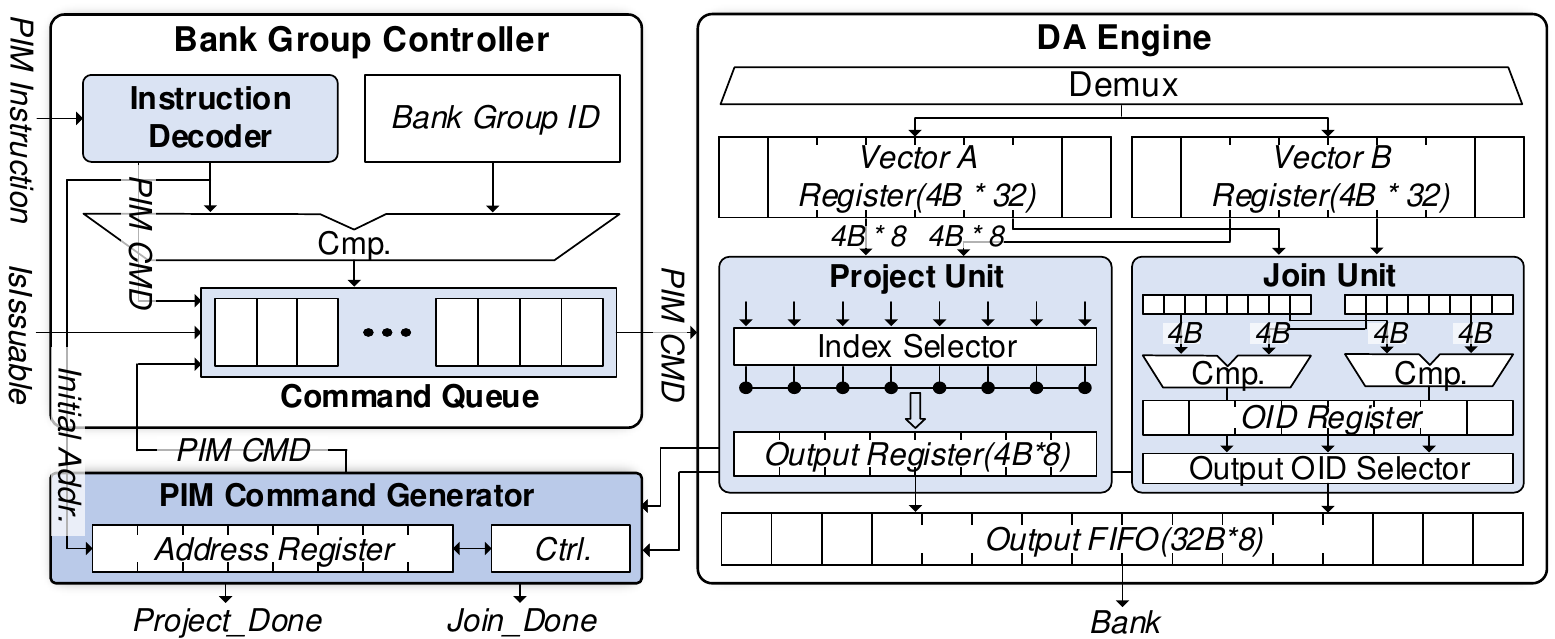}
\caption{BGPU Microarchitecture}
\label{Figure_microarchitecture_BGPU}
\end{figure}

As described in Section~\ref{motivation}, computing condition-oriented operators in memory incurs challenges.To this end, we propose BGPU, as shown in Figure~\ref{Figure_microarchitecture_BGPU}. BGPU leverages the efficiency of the project and join unit the most by optimizing the execution flow and the workload balancing overhead together. It comprises the bank group controller, data analytical engine (DA engine), and PIM command generator.

\textbf{Data Analytical Engine}
The DA engine is composed of two vector registers, the project and join unit, and the output FIFO. For the \textit{project} operator, the OIDs and bitmask are stored in the vector register $A$, and the attribute is stored in the vector register $B$. The project unit can decode either OIDs or bitmask, which indicate the selected tuples in the projected attribute. All the operators except \textit{select} generate a set of OIDs as a result, while the \textit{select} operator generates a simple bitmask, as described in the BPU microarchitecture. Based on the pre-configured addresses and the initial OID values, the project unit first sends the bitmask or OIDs to the PIM command generator, so that it can generate the memory read command for the input attribute, and the memory write command for the output attribute. The index selector of the project unit selects projected tuples among eight 4B tuple data at ${t_{CCDL}}$ period to rate-match the peak bandwidth at the bank group level, assuming DRAM configures 16-bit DQ pins with the burst length of 16. Then, the selected output data are stored in the output register. Depending on the selectivity, the index selector selects less than eight data. If a set of complete eight 4B data are prepared in the output register, it is dispatched to the output FIFO, which eventually goes to the banks. To reduce the read and write turnaround latency, the output FIFO holds on to 256B data and sequentially write them back to the bank. 

For the merge phase of \textit{join}, the two sorted attributes are fetched in the vector register $A$ and $B$, while the two OID sets are stored in the OID registers in the join unit. The join unit receives the two input attributes to merge them by comparing each tuple sequentially, executed by the join controller. The join controller sends addresses for the required attribute data to the PIM command generator which generates the memory read and write command for the next input. In order to rate-match the peak bandwidth of the bank group, it includes two comparators for processing two sets of tuples at a time, or a total of four sets of 4B data and 4B OIDs. The OIDs of the output data which satisfy the join-merge condition are selected by the output OID selector. Then, the OIDs of the two tuples are sent to the output FIFO. Same as the project operator, the output FIFO holds on to a set of output data and sequentially writes them back to the bank.

\textbf{Bank Group Controller}
Instead of sending the intermediate results of the condition-oriented query operators to the host CPU for checking the condition, \sysname supports the bank group controller to take the CPU's role and remove off-chip data movement for host-pim communication. The bank group controller is responsible for managing the PIM commands within bank group. The sources of the PIM commands are either from the PIM instruction or the PIM command generator. The PIM instruction is decoded into PIM commands in the instruction decoder. The PIM commands are generated sequentially and stored in the command queue. On the other hand, the PIM command generator in the bank group generates PIM commands based on the initial configurations received from the BGPU instructions when executing the \textit{project} and \textit{join} operators. When the PIM commands are stored in the queue, it receives $issuable$ signal from the command scheduler and sends out the PIM command.

\textbf{PIM Command Generator}
The PIM command generator conditionally generates PIM commands that are determined by the condition-oriented operations. Since DRAM is a timing deterministic device that the control signal is managed by the strict timing rules, the host memory controller cannot decide when to properly send the next command if the execution flow is non-deterministically decided inside the memory by the bank group controller. This issue can be easily addressed with the simple hand-shaking protocol between the CPU and the PIM device. The host CPU holds the next PIM instruction if the ready signal is not asserted from the device side.

\subsection{Chip-level Command Scheduler}
\sysname includes the PIM command scheduler at the chip level to oversee all the bank group command queues considering the inter-bank timing constraints such as row-to-row activation delay (${t_{RRD}}$) and four-bank activation window (${t_{FAW}}$). Along with the inter-bank timing constraints, the scheduler manages each bank state and counters, which indicates the remaining latency for each command to be issued. It pops the available PIM commands from each bank group command queue. The overhead of the command scheduler is extremely low since the PIM commands generated for data analytics operators have sequential memory access patterns.

\begin{figure}
\centering
\includegraphics[width=\columnwidth]{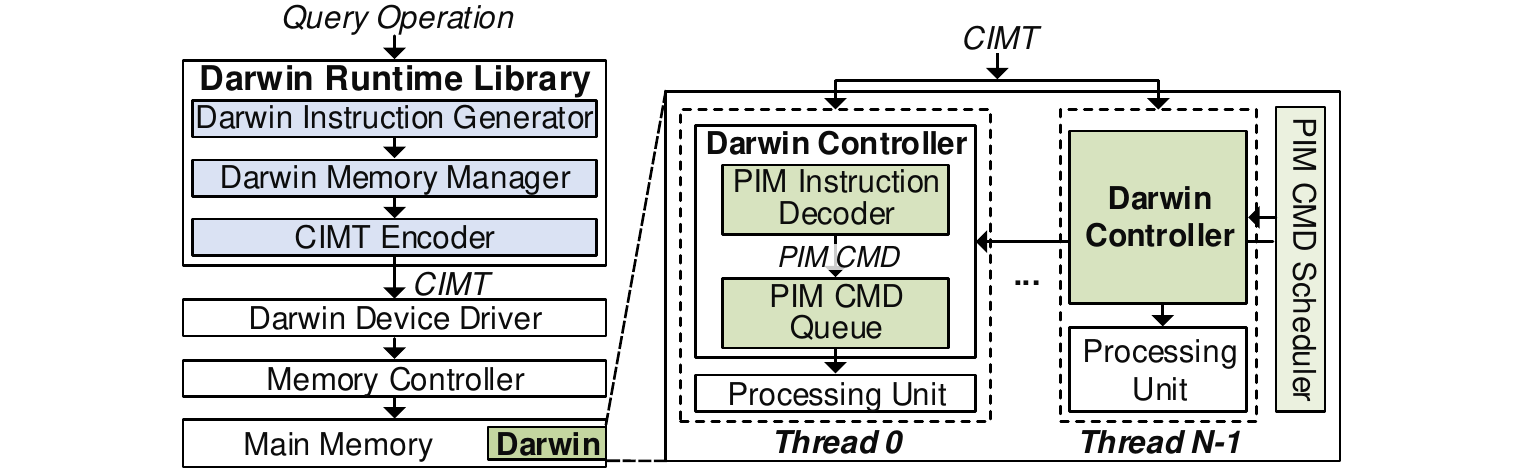}
\caption{\sysname Software Stack}
\label{Figure_Software}
\end{figure}

\subsection{Rank-Level Buffering}
Data movement is unavoidable across the banks, such as workload balancing after the irregular operators, aggregating, projecting, and merging two attributes among the banks. To this end, \sysname supports inter-bank, inter-bank group, and inter-chip communication to move data across the banks efficiently. It supports inter-chip communication by integrating a simple circuit in the buffer chip of the LRDIMM. Figure \ref{Figure_ArchitectureOverview} shows that the instruction decoder, the permute unit, and DQ aligner are only added. When moving data from one chip to another chip, the rank buffer first receives the instruction indicating data read from a bank, waiting ${t_{CAS}}$ to retrieve data from the chip. The PIM instruction also indicates the number of data read in \textit{nCMD}, which generates a sequence of read commands. The rank buffer can receive these data in series and store them in the buffer. Then, the rank buffer receives the instruction indicating data write to a bank in another chip. The instruction includes the permute index which enables re-ordering the data in the rank buffer and writes back to the destination bank. The same process is supported at the chip level and bank group level, which enables inter-bank group and inter-bank communications, respectively. The DQ aligner can change the form of transaction data for the various types of DRAM that \sysname supports to be compliant with the JEDEC's 64-bit DQ pins for the main memory system. It can receive any form of 64-byte data (e.g., 32-bit with 16 burst length) into 64-bit with 8 burst length, which matches the width of the conventional main memory system.

%% file: Contents/6_SoftwareStack.tex
\section{Software Architecture}

\begin{figure}
\centering
\includegraphics[width=\columnwidth]{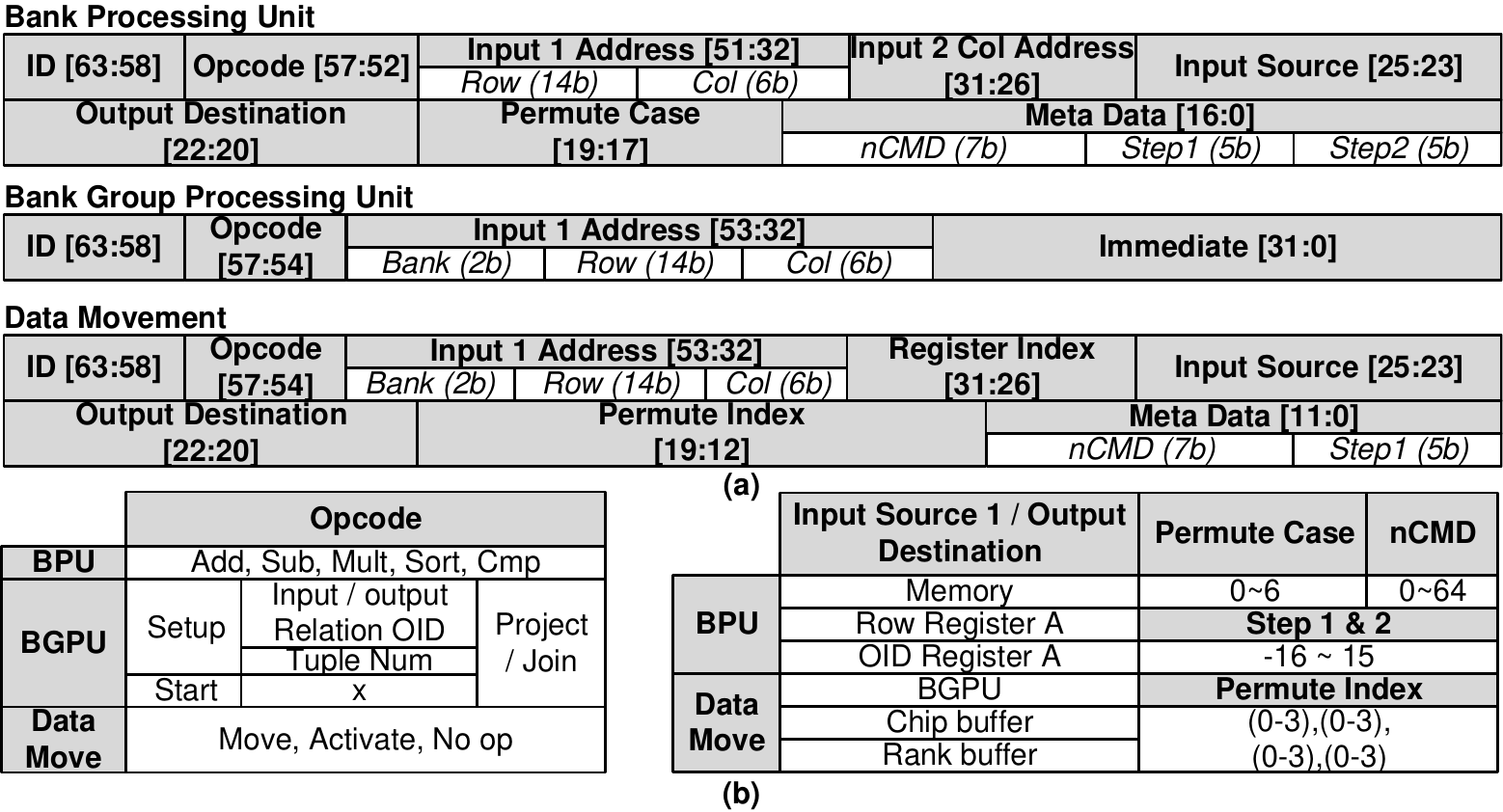}
\caption{Darwin's Instruction Format. (a) PIM (b) Configuration Table}
\label{Figure_Instruction}
\end{figure}

\subsection{Execution Flow of \sysname}
The software stack for \sysname supports the execution with multiple threads, as shown in Figure~\ref{Figure_Software}. The query operation and the corresponding tuples are evenly partitioned into several threads. \sysname runtime library receives query operations from the query plan that are offloaded to \sysname. After receiving query operations, \sysname instruction generator first maps operand data to the memory space in a way that can exploit internal bandwidth the most.
Darwin operates within a separate memory region that is distinct from the memory region utilized by the host, allowing it to bypass coherence issues. This designated area is made uncacheable and enables contiguous mapping of virtual memory addresses to contiguous physical memory addresses. By providing the driver with the starting address of each bank and bank group, the host can map data to a continuous memory space.  Darwin does not engage in virtualization at a lower level than DIMM, such as the rank or bank group level, as it would be both impractical and unnecessary. Since it is typical to use multiple DIMMs to serve a database in data analytics, Darwin supports virtualization at DIMM-level with much flexible control and large memory capacity. Then, it converts query operations into the form of \sysname instruction format, which is shown in Figure~\ref{Figure_Instruction}. \sysname memory manager manages the memory allocated by \sysname device driver for the proper memory addresses. When enough instructions are generated to form a CIMT, the CIMT encoder generates a CIMT, which is then sent to the hardware. In the \sysname hardware, a CIMT instruction is divided into the number of bank groups where each thread is offloaded. Then, each BGPU, including a PIM instruction decoder and a command queue, receives different commands simultaneously. Then, each thread is computed separately on each PU, exploiting the intra-thread and inter-thread data parallelism.

\subsection{Instruction Format}

Figure~\ref{Figure_Instruction} illustrates the 64-bit PIM instruction format. It contains an opcode to determine the type and an ID option to indicate the thread that the instruction is executed on. Depending on the opcode, the instruction format is divided into three categories: BPU, BGPU, and data movement. 

The BPU instruction format and its configurations are shown in Figure~\ref{Figure_Instruction} (a) and (b), respectively. It supports three different input sources (e.g., memory, row register $A$, and OID register $A$) for one input operand of the SIMD unit while the other operand is fixed to the row register $B$. The permute case is used to control the BPU’s permute network. The metadata is used to generate sequential PIM commands using the \textit{nCMD}, \textit{step1}, and \textit{step2}. The \textit{nCMD} determines the number of sequential PIM commands, up to 64, generated by the PIM instruction, while the \textit{step1} and \textit{step2} determine the offset of column addresses for the first and second input sources, respectively. For example, if the column address, \textit{nCMD}, \textit{step1}, and \textit{step2} are configured 0, 4, 1, and 2, respectively, four sequential PIM commands with column addresses of both input operands are configured (0,0), (1,2), (2,4), and (3,6) while the bank and the row addresses are fixed. 

The BGPU instruction format and its options are shown in Figure \ref{Figure_Instruction} (a) and (b). The instructions are mainly for initializing BGPU. The instructions send BGPU with the initial configurations (i.e., input and output OIDs, tuple number, and memory address) to generate PIM commands for computing \textit{project} and \textit{join} operators as well as allocating addresses for intermediate data. After setting the initial configurations, the start instruction initiates the \textit{project} and \textit{join} operators.

The data movement instructions are configured enabling data transfer between the memory and other levels. However, sending PIM instructions for the data movement occupies DQ pins and leaves less room for data transfer. The switching overhead of writing PIM instructions and reading data through the DQ pins becomes even worse if more data transfer occurs. To this end, the \textit{nRD} and \textit{step1} options are enabled for the data movement, generating sequential PIM commands in BGPU and relieving stresses on the DQ pins by the PIM instructions. In addition, the permute index determines the data shuffle order for the permute unit in the rank buffer for inter-chip data movement. 

%\vspace{-0.1in}
\subsection{Data Mapping}
\label{sec_datamapping}
%\vspace{-0.05in}

\begin{figure}
%\vspace{-0.1in}
\centering
\includegraphics[width=\columnwidth]{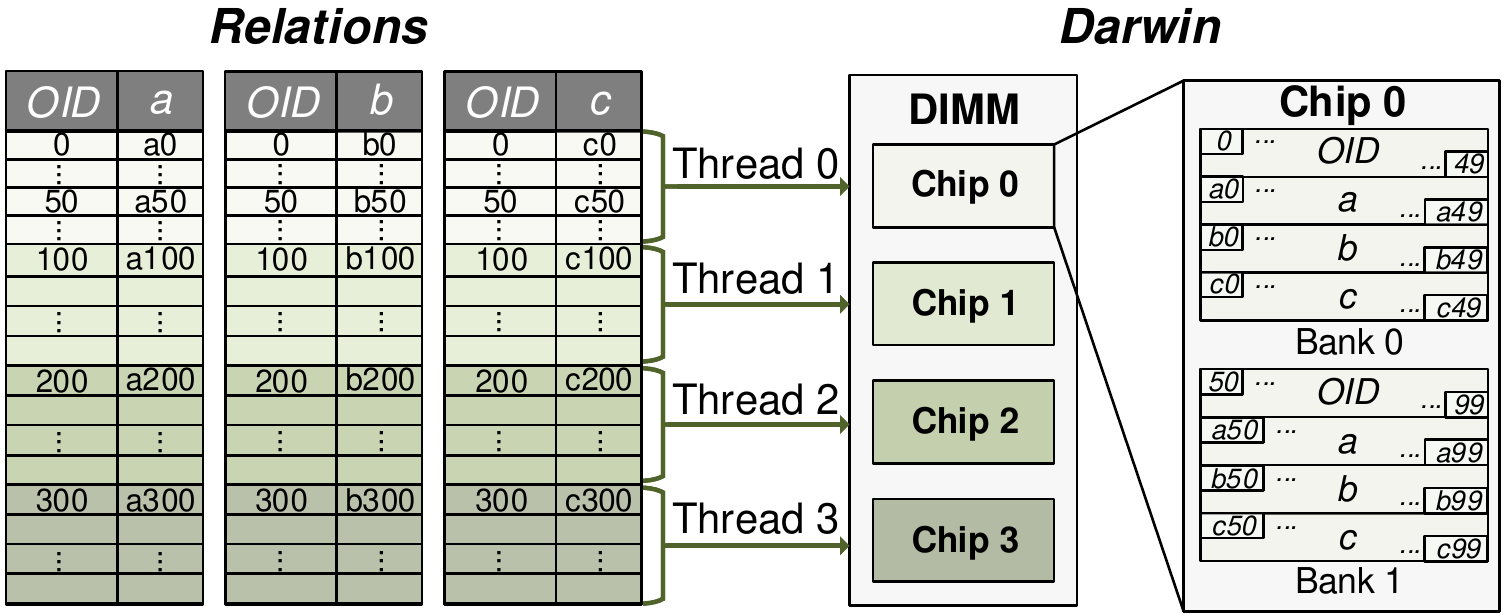}
\vspace{-0.2in}
\caption{Data Mapping}
\label{Figure_Querymapping}
\vspace{-0.1in}
\end{figure}

\sysname adopts the column-oriented DBMS to exploit the data parallelism and DRAM's bandwidth maximally. In the column-oriented DBMS, the attributes are stored separately as array structures to accelerate the analytical query operators, which perform element-wise vector operations on the attributes. In addition, this column-oriented mapping leads to sequential memory access where \sysname can exploit the minimum memory access latency. Furthermore, it adopts relation partitioning to process an analytical query in parallel. Figure \ref{Figure_Querymapping} shows the relation mapping layout of \sysname, assuming DIMM configures 4 chips and 1 bank group per chip with 2 banks per bank group. Since one bank group corresponds to a thread, 4 different threads can be generated. In order to balance the workload across different threads for maximum utilization, columns of attributes and OIDs are evenly partitioned into 4, the total number of threads, then each partition of the attribute is mapped to the corresponding thread. 

%With the same workload across the threads, the partitioned query operators can execute the same amount of computation. 

%% file: Contents/7_Methodology.tex
\section{Methodology}
\subsection{Benchmarks}

\begin{figure}[t]
\centering
\includegraphics[width=\columnwidth]{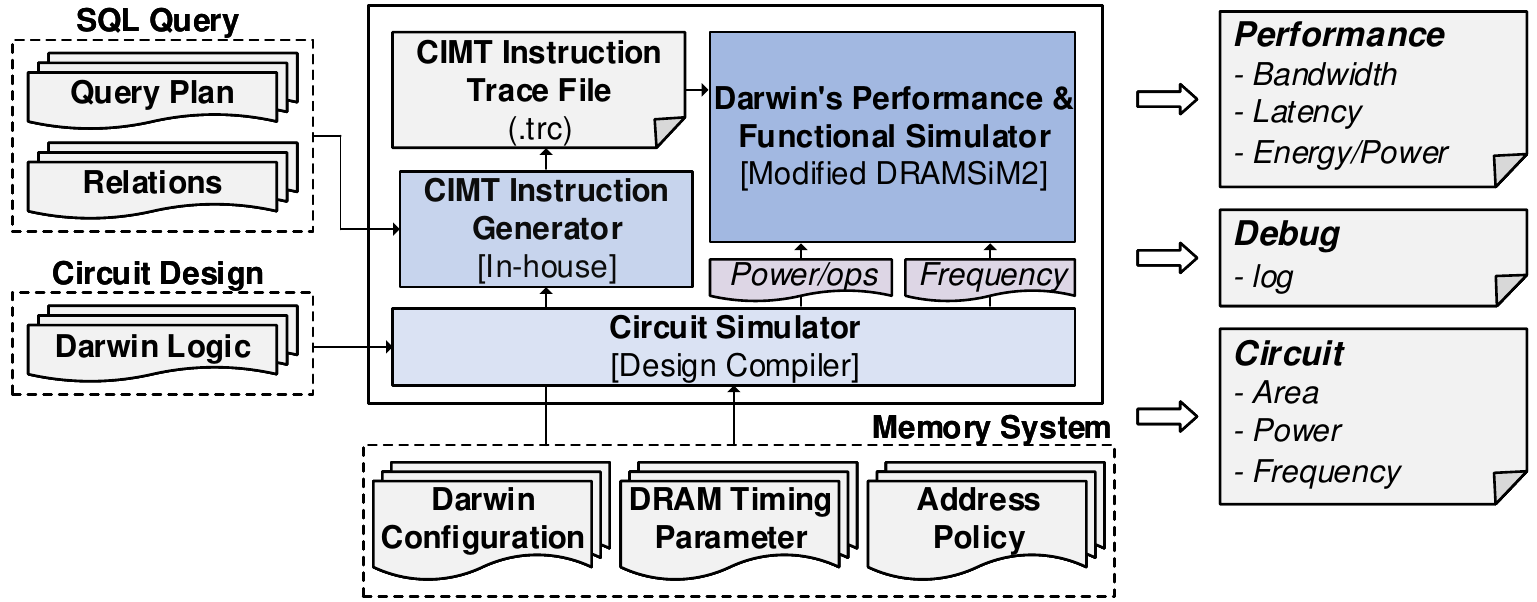}
\caption{\sysname's Simulation Framework}
\label{Figure_framework}
\end{figure}

We use the TPC-H benchmark~\cite{tpc-h} with scaling factor 1 to generate a database whose relations include up to 6,001,215 tuples. For the overall performance of query processing, we evaluate with TPC-H Query 1, which mainly performs \textit{project} and \textit{aggregate} operators on low-cardinality data, and 6, which mainly performs \textit{select} on high-cardinality data. We further evaluate \sysname to see the individual performance of each query operator, such as \textit{select}, \textit{project}, \textit{aggregate}, \textit{sort}, and \textit{join}. For the dataset, we extract 8,388,608 tuples from \textit{lineitem} relation of TPC-H at scaling factor 10. In addition, we use the datasets from Balkensen et al.~\cite{balkesen2013multi} for \textit{join}, where two relations, R and S, have 8,388,608 tuples. Note that, for the dataset for \textit{join}, S is a foreign key to R, which means that every tuple in S has exactly one match to a tuple in R. The dataset assumes a column-oriented model, and both data and OID pairs are 4B integers. 

\subsection{\sysname Simulation Framework}

\textbf{Performance and Functional Simulation}
We modified the DRAMSim2~\cite{rosenfeld2011dramsim2} simulator to support computation of the proposed multi-level functions and CIMT instructions for \sysname, as shown in Figure~\ref{Figure_framework}. MonetDB first converts SQL into an optimized query plan. Receiving a query plan and relations as input, the CIMT instruction generator generates a trace file of CIMT instructions for the operation of \sysname. It also receives the information on the memory system for \sysname configuration and DRAM device parameters to generate addresses and data order of various DRAM devices properly. The instruction trace file is sent to the \sysname's performance and functional simulator. As a result, we can obtain the \sysname's performance results, such as bandwidth utilization and latency, as well as the overall power and energy consumption using the measurement results from the circuit simulator, in executing the query operators. 

\textbf{Area, Power, and Energy Measurement}
The area and power are measured using the Synopsys Design Compiler with a 28nm CMOS technology at 500MHz operating frequency. The power is scaled considering the ${V_{DD}}$ difference. We scale up the area by 80\%~\cite{kim1999assessing} considering the difference between the logic and DRAM process technology, then scale to match the process node. The rank buffer, PIM command scheduler, BGPU, and BPU are synthesized separately, and each logic is scaled with the number of PUs used in the target DRAM device. The energy consumption by the in-memory PUs and internal data movement is measured by the event-driven method that accumulates energy consumption per command to obtain the overall result. The average energy consumption per each PIM command of BPU and BGPU is measured using the PrimePower tool. The power for internal data movement is scaled and modeled from the fine-grained DRAM~\cite{o2017fine}. The energy parameters are integrated into the performance simulator to measure the total energy consumption. 

\begin{table}
\centering
\caption{\sysname Parameters}
\includegraphics[width=\columnwidth]{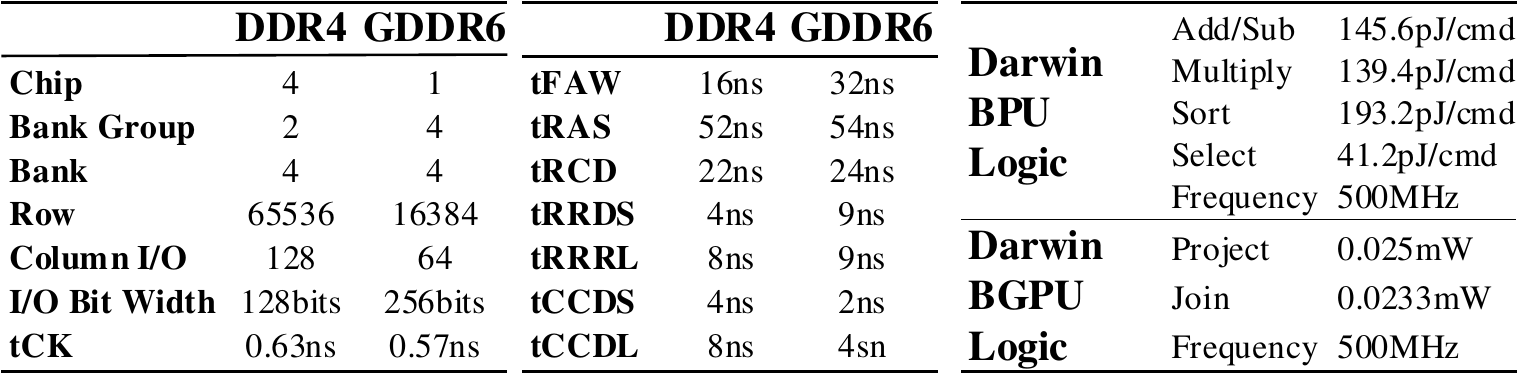}
\label{table_darwin_config}
\end{table}

\subsection{System Configuration}
\textbf{Hardware Baseline}
The TPC-H queries and basic operators are evaluated on four different state-of-the-art architectures: baseline CPU, Mondrian~\cite{drumond2017mondrian}, Newton~\cite{he2020newton}, and \sysname. For the evaluation of PIM architectures, we use the latest GDDR6 configuration for the maximum speedup as the previous research~\cite{lee201425,lee20221ynm} has shown the feasibility of GDDR6-based \sysname in the main memory system for AI applications. The DDR4 configuration is also used for the performance comparison over GDDR6-based \sysname. 

\textbf{CPU}
The baseline CPU is an Intel(R) Xeon(R) Gold 6226R CPU with 512 GB of four channels of DDR4-2933 with a peak bandwidth of 93.84 GB/s. We measure the runtime of TPC-H queries using MonetDB with 64 threads and exclude the query optimization step. For comparison with the basic operators, we implement each operator in C/C++ using the Pthread library~\cite{drepper2003native} to maximize the computation throughput using multiple threads. The sort function in the C++ standard library and hash-join algorithm are evaluated.

\textbf{Previous PIM architectures}
We evaluated the performance of Newton and Mondrian which are representative bank- and rank-level SLPIM architectures, respectively. Newton places PUs at bank-level, while Mondrian integrates PUs at the logic die of HMC. In order to fairly compare \sysname with Newton and Mondrian, we implemented Newton and Mondrian using the same simulation framework as \sysname. We matched the configurations of hardware components for Newton and Mondrian with \sysname to show the benefit of multi-level architecture and CIMT. Both Newton and Mondrian are implemented with the same configurations, such as memory type, frequency and configuration of PU, as \sysname. Newton's microacrhicture of PU is not dedicated for data analytics so we replace Newton's PU with BPU and BGPU at the bank level for the fairness. While Mondrian is SIMD-based architecture that we matched the width of SIMD equal to \sysname. 

\textbf{\sysname}
Table~\ref{table_darwin_config} summarizes the DRAM parameters used for \sysname. We follow typical GDDR6 and DDR4 settings. The GDDR6 has two pseudo-channels (PC) in a package (i.e., chip), where each PC has 16 DQ pins with a burst length of 16. Thus, a total of 64 bytes of data can be transferred per one read or write command. It allows only one package to constitute a rank. To fairly compare the performances of database operators on real CPU hardware and the traced-based simulator, we only compare runtime. The execution time on the baseline CPU is evaluated without query pre-processing steps, such as generating query plan and optimization. For the traced-based simulation, the execution time is evaluated by using a trace file with pre-generated CIMT instructions. In addition, as the simulator always runs optimally, we ensure that the baseline CPU can run optimally by choosing the thread number with the best performance for each operator.

%% file: Contents/8_ExperimentalResult.tex
\begin{figure}[t]
\centering
\includegraphics[width=\columnwidth]{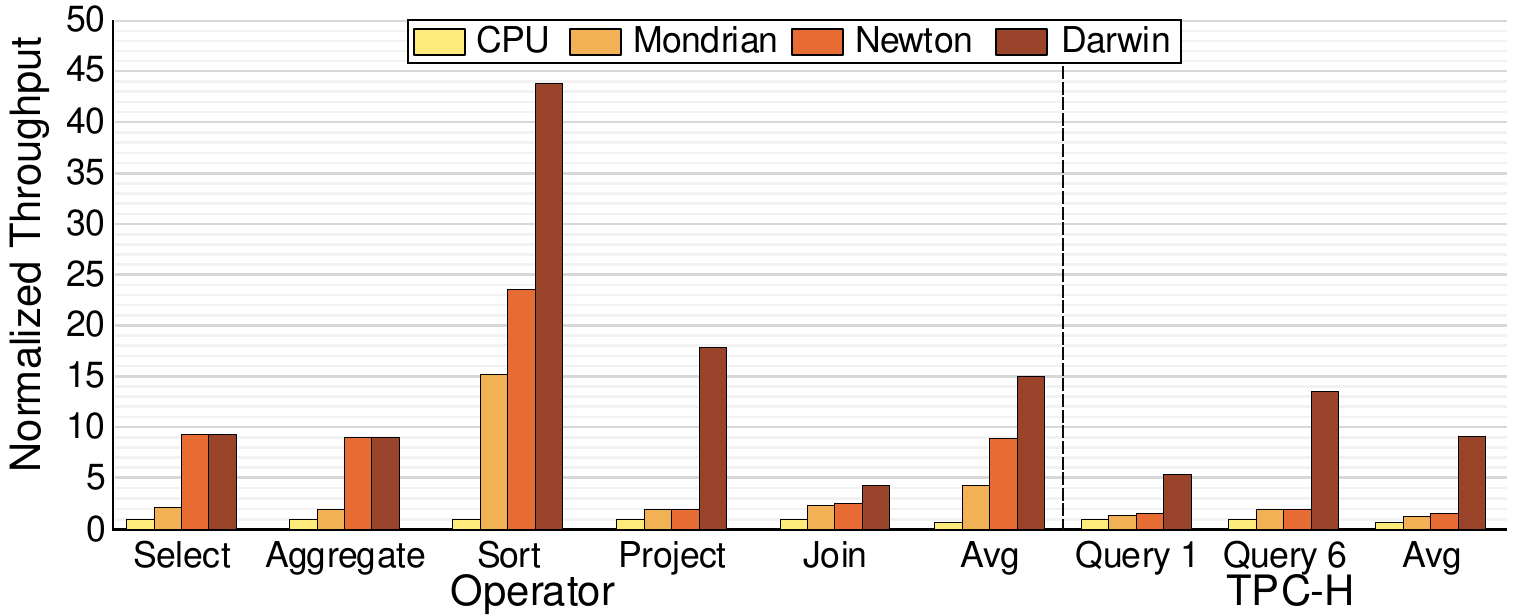}
\caption{Performance Comparison over Baselines}
\label{Figure_graph1}
\end{figure}

\section{Experimental Results}
In this section, we evaluate the benefits of \sysname for various analytics operators and queries. We first compare the performance of \sysname over the state-of-the-art PIM architectures and the baseline CPU and give a detailed analysis of its performance gain on each optimization, bank group-level unit placement, performance comparison over DDR4, and scalability in bank number. Finally, we report the \sysname's physical implementation results and its overhead in GDDR6.
\subsection{\sysname Performance}

\textbf{Comparison to CPU} Figure~\ref{Figure_graph1} shows the speedup of \sysname over the baseline CPU by evaluating the basic query operators. For the comparison over baseline CPU, \sysname achieves 9.3x, 9.0x, 17.8x, 43.9x, and 4.0x faster in the \textit{select}, \textit{aggregate}, \textit{sort}, \textit{project}, and \textit{join}, respectively. The speedup comes from the MLPIM architecture of \sysname that exploits internal parallelism and optimized data movement. The \textit{select}, \textit{aggregate}, and \textit{sort} operators are executed by BPUs utilizing the bank-level parallelism, while the \textit{project} and \textit{join} is executed by BGPUs utilizing the bank group-level parallelism. We further evaluate TPC-H queries for end-to-end query processing. \sysname is 5.4x and 13.5x faster than the baseline CPU in Query 1 and 6, respectively.

\textbf{Comparison to previous PIM Architecture} Mondrain and Newton are evaluated as shown in Figure~\ref{Figure_graph1}. For evaluating the basic query operators and TPC-H queries, both SLPIM architectures show significantly less speedup compared to \sysname. On average, Newton and Mondrian achieve 9.2x and 4.6x higher throughput than the baseline CPU, respectively, while \sysname achieves a much higher throughput of 15.3x. Mondrian shows the least speedup due to the limited bandwidth gain for much further integration of logic. On the other hand, Newton shows no degradation on \textit{select} and \textit{aggregate} operators as compared to \sysname since these operators can be accelerated easily with all bank mode of Newton where all PUs execute on the identical operations. However, \textit{sort}, \textit{project}, and \textit{join} are not simply sped up by having in-memory PUs, requiring further optimization schemes in dataflow with CIMT and multi-level data movement. The performance of Newton and Mondrian are further degraded for the end-to-end TPC-H queries due to a large number of \textit{project} operators on intermediate data and limit the performance gain of only 1.6x and 1.7x, respectively, while \sysname achieves 9.5x higher throughput than the baseline CPU. 

%\vspace{-0.05in}
\subsection{Effect of Optimizations}
%\vspace{-0.05in}

\begin{figure}[t]
\vspace{-0.05in}
\centering
\includegraphics[width=\columnwidth]{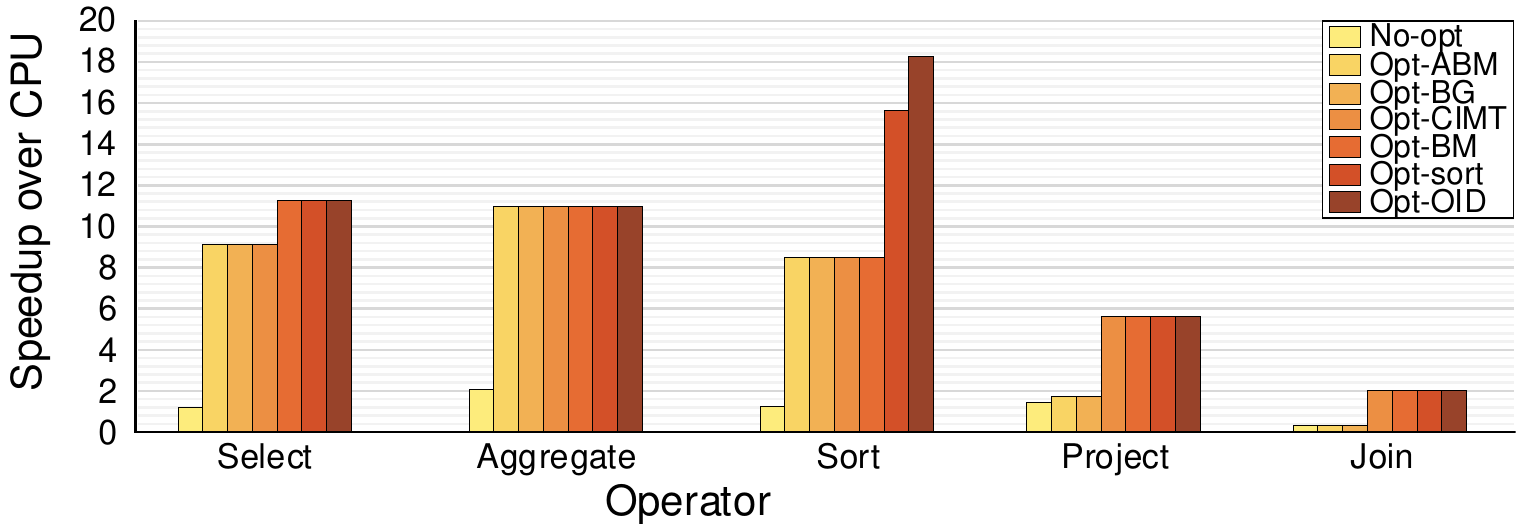}
\vspace{-0.2in}
\caption{Evaluation on Optimization}
\label{Figure_graph2}
\vspace{-0.1in}
\end{figure}

To show the effect of each optimization scheme, we evaluate the speedup when each scheme is applied to the non-optimized \sysname (No-opt), as shown in Figure~\ref{Figure_graph2}. More specifically, No-opt has BPUs that is not optimized for the bitmask register, sort, and CIMT instructions and has the BGPU placed at the rank level. Based on this, we gradually add each optimization: 1) all bank mode (Opt-ABM), 2) BGPU placed at the bank group level (Opt-BG), 3) CIMT (Opt-CIMT), 4) bitmask register (Opt-BM), 5) circuit optimizations for sort (Opt-sort), and 6) OID processing engine (Opt-OID). Although the No-opt has a BPU in each bank exploiting the internal bandwidth, \sysname shows marginal speedup results without any optimizations. It shows only 1.2x, 2.1x, 1.2x, 1.4x, and 0.3x speedup for each operator over the baseline CPU. The performance of \textit{join} operator is degraded significantly because the command bottleneck deters the BPUs in the sort phase and narrow rank bandwidth deters BGPUs in the merge phase.

Each of the schemes significantly improves the performance of \sysname. The CIMT instruction, which applies to all operators, shows the most speedup than the other optimizations by significantly reducing the command bandwidth requirement. The other optimizations are accumulated, and the \sysname with the entire optimizations achieves 9.2x, 5.2x, 14.7x, 3.3x, and 5.8x speedup for each operator compared to the No-opt version, which shows 11.2x, 10.9x, 18.1x, 5.6x, and 2.0x higher performance than the baseline CPU.

\textbf{Optimization on BPU}
The effect of the all bank mode and PIM instruction on BPU are negligible. The reason is that the BPU receives the same instruction for each of the regular operators to leverage the data parallelism. The BPU is rather optimized to reduce computation for the speedup. First, bitmask is used instead of OID for the \textit{select} operator where the memory footprint for the output is reduced by 32x. In addition, the throughput of the \textit{select} with the bitmask (Opt-BM) increases by 11.2x than the No-opt due to the reduced number of writing the output. Second, the compute circuit for the \textit{sort} is optimized to reduce the computation overhead. The optimizations on the sort logic (i.e., SIMD and permute units, Opt-sort) and OPE (opt-OID) reduce the number of compute commands for the bitonic sort by 4x and 2x, respectively. The Opt-sort and Opt-OID achieve 15.5x and 18.1x higher throughput than the No-opt, respectively.

\begin{figure}[t]
\centering
\includegraphics[width=\columnwidth]{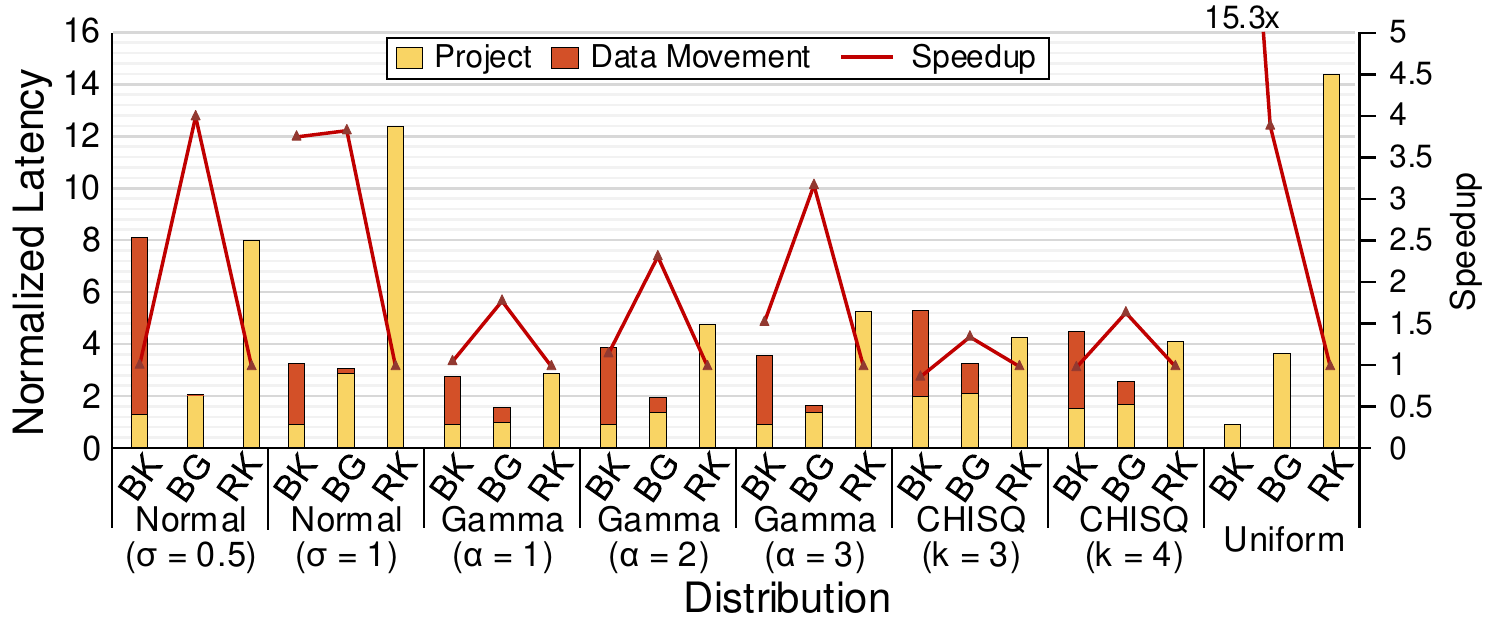}
\vspace{-0.1in}
\caption{Evaluation on BGPU Placement}
\label{Figure_graph3}
\vspace{-0.1in}
\end{figure}

\textbf{Benefit of PIM Instruction}
In data analytics, the command bandwidth is critical as the performance can be bounded by the command bottleneck. The \textit{select}, \textit{aggregate}, and \textit{sort} operators, where the BPU can perform with massive data parallelism, the conventional DRAM command scheme can not provide any speedup to \sysname since each bank receives a command separately. Both all bank mode and PIM instruction show benefit since all BPUs can receive the same command due to the data parallelism. However, in the \textit{project} and \textit{join} operators, simply moving the BGPU from the rank level (Opt-ABM) to the bank group level (Opt-BG) does not provide any speedup with the all bank mode. Each BGPU at the bank group level requires different commands. Thus, the all bank mode cannot leverage the benefit of bank group-level processing. CIMT instruction, which gives different command threads at each bank group level, can effectively leverage the benefit of \sysname. As a result, the CIMT instruction achieves 2.3x and 3.6x speedup in the \textit{project} and \textit{join} operations, respectively. 

%\vspace{-0.06in}
\subsection{Internal Data Movement of MLPIM }
%\vspace{-0.05in}

To see the benefit of exploiting BGPU at the bank group level over the bank and rank levels, we further evaluate the performances of project units depending on its location. Figure \ref{Figure_graph3} shows the evaluation results of BGPU at the different levels performing the \textit{project} operator and workload balancing on eight different data distributions (i.e., normal, gamma, chi-squared and uniform distributions). The \textit{project} operator is performed at each memory level, and the additional data movement is applied to evenly distribute the output data, which will later be fed as input to the following operator. The latency reduces as the placement of BGPU goes lower to the bank level, but the data movement latency increases since the \textit{project} operation is only executed within the level. In the uniform distribution, there is no additional data movement overhead since the workload is perfectly balanced. Therefore, there is only the latency for the execution of the \textit{project} operator, achieving the highest speedup in the bank level. However, the uniform distribution is an ideal case that is unlikely to occur. Thus, with the normal, gamma, and chi-squared distributions, the BGPU in the bank group level shows the highest speedup of average 4.0x than the rank level.

\begin{figure}[t]
\centering
%\vspace{-0.05in}
\includegraphics[width=\columnwidth]{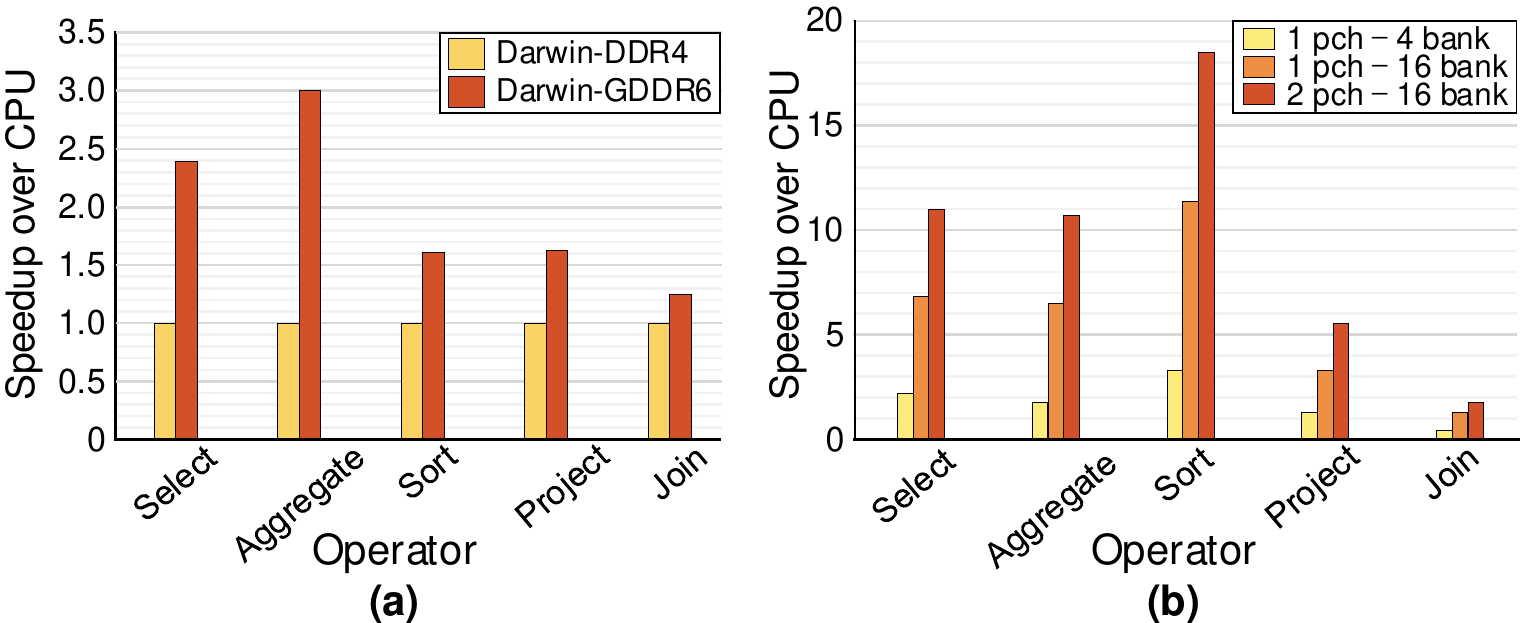}
\vspace{-0.2in}
\caption{\sysname Scalability. (a) DDR4 Comparison (b) Memory Configuration}
\vspace{-0.1in}
\label{Figure_graph4}
\end{figure}

%\vspace{-0.05in}
\subsection{Comparison over DDR4}
%\vspace{-0.05in}
Figure~\ref{Figure_graph4} (a) shows the normalized throughput of DDR4 based \sysname (\sysname-DDR4) and GDDR6 based \sysname (\sysname-GDDR6). The memory configurations are shown in Table~\ref{table_darwin_config}. A single chip of GDDR6 is compared with four chips of DDR4, forming a rank, to match the number of banks, BPUs, and BGPUs between them. The capacity of \sysname-DDR4 is 4 times larger than the \sysname-GDDR6. However, \sysname-GDDR6 achieves up to 3.0x higher throughput than \sysname-DDR4 since GDDR6 provides about two times faster column access and two time wider I/O bit width than DDR4.

%\vspace{-0.05in}
\subsection{\sysname Scalability}
%\vspace{-0.05in}
Figure~\ref{Figure_graph4} (b) shows the speedup of \sysname for the basic query operators as we vary its memory configuration among 1 PC with 4 banks, 1 PC with 16 banks, and 2 PC with 16 banks. The internal bandwidth and the PUs increase linearly as the number of banks increases. However, the speedup dampens as the number of banks increases due to the increased amount of internal data movement. The average speedup of \sysname when the bank number increases from 4 to 16 with 1 PC is 3.2x. On the other hand, the average speedup when increasing the PC from 1 to 2 with 16 banks is 1.6x.

%\vspace{-0.05in}
\subsection{Area, Frequency, Power, and Energy}
%\vspace{-0.05in}

The areas of BPU, BGPU, and rank buffer are measured 0.104${mm^2}$, 0.043${mm^2}$, and 0.078${mm^2}$. Once scaled and summated, the total area overhead is measured 3.752${mm^2}$ which is only 5.6\% of the GDDR6's die area~\cite{kim201816}.

The energy consumption is evaluated as shown in Figure~\ref{Figure_graph5}. In the ideal CPU, the energy consumption is measured only by the data movement between the CPU and the memory. Since we assume the CPU is ideal with unlimited computation capacity and speed, it does not take any delay or energy for any operator execution.
%We assume that the host CPU with unlimited computation capability performs each operator, meaning that there is no energy consumption and delay by the CPU execution. 
This guarantees that the peak bandwidth is utilized, and there is no redundant read or write on the same data address. The background energy is increased in \sysname due to the longer execution time than the ideal case, however, due to the reduced off-chip movement, the overall energy drops. As a result, the reduced I/O movement brings all operators to achieve energy savings significantly. \sysname reduces its energy consumption by 45.4\%, 10.2\%, 67.4\%, 47.1\%, and 84.7\% in \textit{join, sort, project, aggregate,} and \textit{select}, respectively.

%% file: Contents/9_RelatedWork.tex
%\vspace{-0.1in}
\section{Related Work}
%\vspace{-0.05in}
\textbf{PIM and NMP} Newton, HBM-PIM, TransPIM, McDRAM, Ambit, and SIMDRAM~\cite{he2020newton,lee2021hardware,zhou2022transpim, shin2018mcdram, seshadri2017ambit, hajinazar2021simdram} support the regular or non-condition-oriented workloads to avoid data dependent dataflow by accelerating memory-bound vector operations exploiting internal parallelism
of DRAM. \cite{park2021trim, kwon2019tensordimm,ke2020recnmp} accelerates a recommendation system where the gather-and-scatter operations are the main target.

\textbf{Accelerating Data Analytics}
On-chip accelerator Q100~\cite{wu2014q100} exploits pipeline in query processing with heterogeneous processing cores to minimize the memory access. Mondrian and Polynesia\cite{drumond2017mondrian,boroumand2021polynesia} integrate circuits in logic die of 3D stacked memory, which is much further from the DRAM cells and loses the internal bandwidth. To the best of our knowledge, \sysname is the first proposal that accelerates data analytics operators reducing the overhead by reusing the hierarchical structure of the main memory system.

\section{Discussion}
\textbf{Applying \sysname to other workloads} Although \sysname is designed to target data analytics, it can still accelerate other similar workloads that have large memory footprint, memory bounded, and especially having wide data dependency. Deep neural network (DNN) workloads, such as transformer and LSTM, are applicable to Darwin. Its memory-bounded algebra operations, such as matrix vector multiplication and matrix-matrix multiplication, can be easily accelerated with BPU's SIMD. Furthermore, multi-level characteristic of Darwin can accelerate the frequent output and partial sum data movement of these workloads where they are incurred from data partitioning among the memory nodes due to the large sized weight. Especially, \sysname is applicable to accelerate sparse matrix multiplication with its efficient in-memory network to gather non-zero elements and perform multiplication in BPU. In addition, Darwin can accelerate recommendation system where it includes gather-and-reduction and fully-connected workloads. Darwin can gather data from wide range of memory space in parallel using BGPU, while fully-connected layers are accelerated using BPU's SIMD.

\begin{figure}[t]
\centering
%\vspace{-0.05in}
\includegraphics[width=\columnwidth]{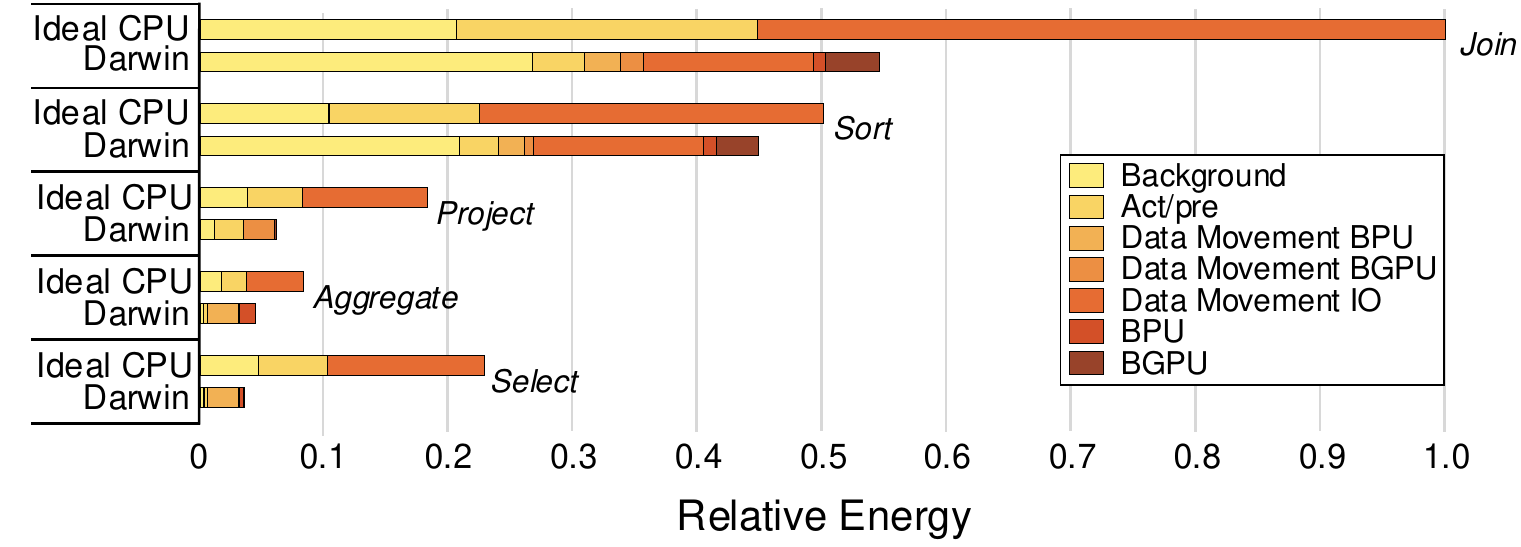}
\vspace{-0.2in}
\caption{Evaluation on Relative Energy Consumption}
\label{Figure_graph5}
\vspace{-0.1in}
\end{figure}

%% file: Contents/10_Conclusion.tex
%\vspace{-0.1in}
\section{Conclusion}
%\vspace{-0.05in}
We propose \sysname, a practical LRDIMM-based multi-level PIM architecture for data analytics. We addressed the issues in adopting PIM for data analytics through the three contributions. First, \sysname reduces overhead of integrating additional logic in DRAM by reusing the conventional DRAM architecture that fully exploits the multi-level of DRAM, while maximizing the internal bandwidth. Second, \sysname places the BGPU to mitigate the data movement overhead and load balancing across the banks, while performing the condition-oriented memory-bounded data analytics operators. Third, CIMT instruction is adopted to address the command bottleneck to enable separate control of multiple PUs simultaneously. The simulation results on the major five data analytics operators, such as \textit{select}, \textit{aggregate}, \textit{sort}, \textit{project}, and \textit{join}, show that the GDDR6 based \sysname achieves up to 43.9x over the baseline CPU. \sysname is more energy-efficient than the ideal CPU systems by 85.7\%, while the additional area overhead is only 5.6\%.